\documentclass[
amsmath,
amssymb,
aps, 
pra,
twocolumn, 
groupedaddress,
nobalancelastpage,
floatfix]
{revtex4-2}
\usepackage{mathtools}
\usepackage[colorlinks]{hyperref}
\usepackage{lipsum}
\usepackage{bbold}
\usepackage{bm}
\usepackage{dsfont}
\usepackage{enumerate}
\usepackage{graphicx}
\usepackage{float}
\usepackage {xcolor}
\usepackage{cleveref}
\usepackage{amsmath}

\usepackage{lipsum}

\renewcommand{\vec}[1]{\bm{#1}}	

\DeclarePairedDelimiterX{\mean}[1]{\langle}{\rangle}{
	{#1}
}
\DeclarePairedDelimiterX{\abs}[1]{\lvert}{\rvert}{
	{#1}
}
\DeclarePairedDelimiterX{\norm}[1]{\lVert}{\rVert}{
	{#1}
}
\DeclarePairedDelimiterX{\bra}[1]{\langle}{\rvert}{#1}

\DeclarePairedDelimiterX{\ket}[1]{\lvert}{\rangle}{#1}

\DeclarePairedDelimiterX{\mel}[3]{\langle}{\rangle}{
	{#1}\delimsize\vert {#2}\delimsize\vert {#3}
}
\DeclarePairedDelimiterX{\inner}[2]{\langle}{\rangle}{
	{#1} \delimsize\vert{#2}
}
\DeclarePairedDelimiterX{\dyad}[2]{\lvert}{\vert}{
	{#1} \delimsize\rangle \delimsize \langle{#2}
}

\begin{document}
	\title{High-Order Harmonic Generation with Beyond-Semiclassical Emitter Dynamics: A Strong-Field Quantum Optical Heisenberg Picture Approach}
	
	\author{Christian Saugbjerg Lange}			
	\author{Ella Elisabeth Lassen}						
	\author{Rasmus Vesterager Gothelf}		
	\author{Lars Bojer Madsen}					
	\affiliation{Department of Physics and Astronomy, Aarhus University, Ny Munkegade 120, DK-8000 Aarhus C, Denmark}
	\date{\today}
	\begin{abstract}
	Quantum-optical descriptions of strong-field processes have attracted significant attention in recent years. Typically, the theoretical modeling has been conducted in the Schrödinger picture, where results are only obtainable under certain approximations, while, in contrast, the Heisenberg picture has remained relatively unexplored. In this work, we develop an accurately controlled perturbative expansion of the time-evolution operator in the Heisenberg picture and derive beyond-semiclassical corrections to the emitter dynamics due to the coupling to the quantized electromagnetic field, capturing effects of the quantum fluctuations present in the latter. We focus on high-order harmonic generation (HHG), where the approach is accurate in parameter regimes of current interest and it gives closed-form expressions for key observables. This formulation not only simplifies numerical calculations compared to the Schrödinger-picture approach but also provides a clear correspondence between nonclassical features of the emitted light and the underlying induced dynamics of the generating medium including quantum fluctuations. Moreover, the Heisenberg framework naturally yields scaling relations with the number of independent emitters, enabling us to assess whether nonclassical behavior should persist under typical experimental conditions involving large emitter ensembles. Interestingly, we find that the degree of squeezing increases with the number of emitters, whereas the photon statistics approaches a classical Poissonian distribution in the many-emitter limit. We also find that the beyond-semiclassical emitter dynamics significantly enhances the degree of squeezing of the emitted light. Our work advances the theoretical understanding of quantum-optical HHG and introduces an accessible and well-controlled framework to describe realistic experiments.
\end{abstract}

\maketitle


\section{Introduction} \label{Sec:Introduction}
Strong-field physics, the branch of science that deals with intense light-matter interactions, has for decades successfully described experimental results for various processes such as above-threshold ionization (ATI) and high-order harmonic generation (HHG) using semiclassical theory. In such a description, the electronic medium is treated quantum mechanically while the involved light fields are treated classically. Consequently, such a description does not incorporate the quantum nature of the light fields, forgoing the full quantum-mechanical description of the strong-field-induced dynamics. Treating the involved light fields quantum mechanically has, in recent years, sparked a growing interest in the emerging research area of strong-field quantum optics, which has already led to novel findings, expanding the well-known and successful results predicted by a semiclassical theory. Adopting a quantum optical description enables one to improve upon the semiclassical theory by investigating how the quantized nature of the electron is transferred to the quantized nature of the light fields. 

Already, a quantum optical description has been used in different systems and processes. For instance, free electrons interacting with quantized light are now studied, revealing how the photon statistics of this driving field modifies the electron dynamics, which can be used in quantum information science \cite{Ruimy2025} or as a probe for the driving field \cite{Giulio2019, Kfir2021, BenHayun2021, Dahan2021, Dahan2023, Gorlach2024}. Additionally, electron spectra from the strong-field process of ATI reveal how the quantum nature of the driving field modifies the tunneling process when compared to a classical coherent driver \cite{FangLiu2023, Liu2025}. Similarly, studies of HHG clearly show both theoretically \cite{Gorlach2023, EvenTzur2023, EvenTzur2024, Gothelf2025} and experimentally \cite{Rasputnyi2024, Lemieux2025, EvenTzur2025}, that the harmonic spectra are drastically changed when the electronic system is driven by nonclassical driving fields. 

So far, the strong-field process that has received the most attention from a quantum optical perspective is HHG driven by an intense coherent laser. By driving the system with a classical laser, the observed nonclassicallity of the generated light is purely due to the electronic response. Hence, the emitted nonclassical light can serve as a probe for the electron dynamics, which widens the use of HHG as a spectroscopic tool, while at the same time it invites one to design electron systems that generate a strong nonclassical response, such that the light might have usage in related research fields such as quantum sensing \cite{Lewenstein2024AttoAndQI}. Already, the quantum-optical response from many different electron systems driven by a classical coherent field have been studied, including atomic gasses \cite{Gorlach2020, Lewenstein2021,Stammer2023, Rivera-Dean2024e, DeLaPena20225, Stammer2024a}, molecules \cite{Rivera-Dean2024d}, solids \cite{Rivera-Dean2024f}, entangled many-body systems \cite{Pizzi2023}, and strongly correlated materials \cite{Lange2024a, Lange2025a} including the effects of an exciton in a Mott insulator \cite{Lange2025b}. 

Historically, strong-field processes have mostly been described using the Schrödinger picture and semiclassical theory. On the other hand, studies in quantum optics have often favored the Heisenberg description of the photonic operators.
So far, building from semiclassical experience, most studies of quantum optical HHG have used the Schrödinger picture, yielding a set of coupled equations of motion for the involved photonic states driven by light-matter correlations. Unfortunately, the Hilbert space for the photonic degrees of freedom capturing all possible numbers of photons in all relevant modes is exponentially large \cite{Lange2025a}, impeding an exact numerical treatment. Consequently, to proceed, approximations have been made. This includes, e.g., a decoupling of the photonic modes, i.e., neglecting all dynamical correlations between different photonic frequencies and polarizations \cite{Gorlach2020, Lange2024a, Lange2025a, Lange2025b}, or keeping only two photonic states, known as a two-channel approximation, which yields a perturbative solution \cite{Yi2024, Rivera-Dean2024e, Rivera-Dean2024d, Rivera-Dean2024f}, or simply neglecting the time correlations of the driven dipole \cite{Lewenstein2021, Rivera-Dean2022, Stammer2023, Bhattacharya2023, Stammer2024, Stammer2025b}. The consequence of the latter approximation is well understood: Neglecting time correlations of the induced dipole, i.e., assuming that the expectation value of the induced dipole is uncorrelated to its past, the HHG process always generates a coherent state. This approximation, hence, prohibits the generation of quantum light from any electronic system. The two former approximations, in contrast, in general predict a nonclassical response in the emitted HHG. However, as exact results are not obtainable, the approximations, though justified on physical intuition, remain without benchmark, and an alternative description is needed to qualify these approximations. A formulation of strong-field quantum optics in the Heisenberg picture is a natural candidate for such an alternative description, building on decades of experience from work on quantum optics. Early works have considered HHG using the Heisenberg picture \cite{Sundaram1990}, however, only calculating the harmonic spectrum, while very recent work has also considered photon statistics of the emitted light \cite{Stammer2025}.

In this work, we derive a description of quantum optical HHG using the Heisenberg picture, which for the first time explicitly captures beyond-semiclassical dynamics of the generating emitter. As the exact dynamics are challenging to obtain, we expand a transformed time-evolution operator in orders of the resulting (weak) light-matter coupling, yielding closed-form expressions for the observables of interest. In this way, we approximate the dynamics in a consistent and controlled manner, which serves as a mirror to the Schrödinger-picture description with related approximations, qualifying the latter. This description, which utilizes the Heisenberg picture, comes with a number of advantages when compared to approaches using the Schrödinger picture. The theory allows us to conclude that the higher-order correction terms in the light-matter coupling constant are negligible in the parameter regime of current HHG experiments, i.e., the theory is accurate. Moreover (i) electron dynamics include quantum-induced corrections to semiclassical dynamics, which (ii) affects the considered photonic observables. (iii) The scaling with the number of identical and independent emitters follows naturally for all photonic observables. (iv) The derived expressions clearly link the photonic observable to the underlying electron dynamics. (v) The numerical cost is smaller compared to the approach in the Schrödinger picture. With the established theory, we calculate the HHG spectrum, the degree of squeezing, and the photon statistics from both an ensemble of atoms and a strongly correlated material modeled by the Fermi-Hubbard Hamiltonian, showcasing the generality of the derived theory. We find a good agreement with the results obtained via the Schrödinger picture, qualifying the employed approximations in the latter. 

The paper is organized as follows. In Sec. \ref{Sec:Theory}, we show the exact dynamics in the Heisenberg picture and discuss how a numerical evaluation is prohibited due to the lack of a closed-form expression. This is followed by a presentation of a controlled and perturbative expansion of the time-evolution operator of the emitter dynamics, yielding closed-form expressions for the photonic observables of interest. In Sec. \ref{Sec:Results}, the numerical results are presented and discussed for both an atomic system and a strongly correlated material. In Sec. \ref{Sec:Discussion}, we compare the established approach derived in the Heisenberg picture to its counterpart in the Schrödinger picture and discuss the need to include quantum corrections to semiclassical emitter dynamics, followed by a discussion of the experimentally expected degree of squeezing and the role of the so-called transition dipoles. Finally, the conclusions and an outlook are given in Sec. \ref{Sec:Conlcusion}. The appendices App. \ref{App:Photon_statistics}-\ref{App:product_ansatz_consequence} provide detailed calculations that support the main text.

\section{Theory} \label{Sec:Theory}
We consider a system of $N$ independent dipoles all driven by the same intense laser field described by coherent states. The Hamiltonian for such a system is given as (atomic units used throughout)
\begin{equation}
	\hat{H}(t) =  \sum_{j=1}^{N} \left[\dfrac{(\hat{\vec{p}}_j + \hat{\vec{A}})^2}{2} + \hat{U}_j \right] + \hat{H}_F, \label{eq:H_general_1} 
\end{equation}
where $\hat{p}_j$ is the momentum operator and $\hat{U}_j$ describes the potential for the $j$'th emitter, respectively, $\hat{H}_F = \Sigma_{\bm{k}, \sigma} \omega_k \hat{a}_{\bm{k}, \sigma}^\dagger \hat{a}_{\bm{k}, \sigma}$ is the Hamiltonian of the free field with $\hat{a}_{\bm{k}, \sigma}$ ($\hat{a}_{\bm{k}, \sigma}^\dagger)$ being the photonic annihilation (creation) operator for the mode with wave vector $\bm{k}$, frequency $\omega_k = c \lvert \bm{k} \rvert$ and polarization $\sigma$. The quantized vector potential entering Eq. (\ref{eq:H_general_1}) is given by 
\begin{equation}
	\hat{\vec{A}} = \sum_{\bm{k}, \sigma} \dfrac{g_0}{\sqrt{\omega_k}} (\hat{a}_{\bm{k}, \sigma} + \hat{a}_{\bm{k}, \sigma}^\dagger) \hat{\bm{e}}_\sigma, \label{eq:A_operator_1}
\end{equation}
in the dipole approximation where $g_0 = \sqrt{2 \pi/V}$ is the coupling constant with quantization volume $V$ and polarization unit vector $\hat{\bm{e}}_\sigma$ assumed to be real. We note that the dipole approximation is expected to be exact for the characterized fields \cite{Gorlach2020}.

We emphasize that each independent emitter might consist of a multielectron system. For instance, if we consider an ensemble of independent atoms, each momentum operator is the sum of single-electron operators for a specific atom, i.e., $\hat{\bm{p}}_j = \sum_l \hat{\bm{p}}_j^{(l)}$, where $\hat{\bm{p}}_j^{(l)}$ is the momentum operator for the $l$'th electron on atom $j$, and $\hat{U}_j$ similarly describes the potential for all the electrons of atom $j$. In this work, we consider only independent and identical emitters, treating interacting systems as a single instance of an independent system.

The initial condition for the combined state of the system is
\begin{equation}
	\ket{\Psi(0)} =  \otimes_{j=1}^N \ket{\phi^{(j)}_i} \otimes_{\{(\bm{k}_L, \sigma_L)\}} \ket{\alpha_{\bm{k}_L, \sigma_L}}  \otimes_{\bm{k}, \sigma \notin \{(\bm{k}_L, \sigma_L)\}} \ket{0_{\bm{k}, \sigma}}, \label{eq:initial_state_exact}
\end{equation}
that is, the emitters are in a product state with emitter $j$ in its initial state $i$, denoted as $\ket{\phi_i^{(j)}}$, and the modes present in the driving laser, $\{(\bm{k}_L, \sigma_L)\}$, are in coherent states, while all other photonic modes are in the vacuum state. Equations (\ref{eq:H_general_1}) and (\ref{eq:initial_state_exact}) are the starting points for the following subsections.

\subsection{Exact dynamics} \label{Sec:exact_dynamics}

We calculate the dynamics of the system via the Heisenberg equation of motion for the photonic operator $\hat{a}_{\bm{k}, \sigma}$ using Eq. (\ref{eq:H_general_1})
\begin{equation}
	\dfrac{d}{dt} \hat{a}_{\bm{k}, \sigma}(t) = - i \omega_k \hat{a}_{\bm{k}, \sigma}(t) - i \dfrac{g_0}{\sqrt{\omega}_k} \hat{\bm{e}}_\sigma \cdot \hat{\bm{Q}}(t), \label{eq:Heisenberg_EOM_exact}
\end{equation}
where we have defined $\hat{\bm{Q}}(t) = \hat{\mathcal{U}}^\dagger(t) \hat{\bm{Q}} \hat{\mathcal{U}}(t)$ with $\hat{\bm{Q}}= \sum_{j=1}^N \hat{\bm{p}}_j$ being the sum of the momentum operators of the independent emitters and $\hat{\mathcal{U}}(t)$ the exact time-evolution operator for the joint light-matter system. Note that upon obtaining Eq. (\ref{eq:Heisenberg_EOM_exact}), we have neglected the term proportional to $\hat{\bm{A}}^2$, see, e.g., Refs. \cite{Faisal1973,Gorlach2020, Lange2024a, LangeThesis} for a discussion. The integral form of Eq. (\ref{eq:Heisenberg_EOM_exact}) is given as
\begin{equation}
	\hat{a}_{\bm{k}, \sigma}(t) = \hat{a}_{\bm{k}, \sigma}(0) e^{-i \omega_k t} - i \dfrac{g_0}{\sqrt{\omega_k}} \int_{0}^{t} dt' \hat{\bm{e}}_\sigma \cdot \hat{\bm{Q}}(t') e^{-i\omega_k(t-t')}. \label{eq:Heisenberg_solution_exact}
\end{equation}	
From Eq. (\ref{eq:Heisenberg_solution_exact}), one can construct photonic observables of interest. In this work, we focus on the harmonic spectrum, the degree of squeezing, and photon statistics quantified by the second-order correlation function $g_{\bm{k}, \sigma}^{(2)}(0)$. In deriving expressions for these observables, we make two important approximations. First, we assume that all emitters are independent, i.e., noninteracting. Secondly, we assume that all emitters are identical, i.e., that they are the same type of electron system experiencing the same driving field. Mathematically, these assumptions are described as
\begin{subequations} \label{eq:independent_and_identical_emitters}
	\begin{equation}
		\langle \hat{\bm{e}}_\sigma \cdot \hat{\bm{p}}_i(t) ~ \hat{\bm{e}}_\sigma \cdot \hat{\bm{p}}_j(t) \rangle = \langle \hat{\bm{e}}_\sigma \cdot \hat{\bm{p}}_i(t) \rangle \langle \hat{\bm{e}}_\sigma \cdot \hat{\bm{p}}_j(t) \rangle, \quad (i \neq j)
	\end{equation}
	\begin{equation}
		\langle \hat{\bm{p}}(t) \rangle = \langle \hat{\bm{p}}_j(t) \rangle  \quad \text{for all $j$},
	\end{equation}
\end{subequations}
where the latter also extends to expectation values of all higher moments. In Eq. (\ref{eq:independent_and_identical_emitters}), we have denoted $\hat{\bm{p}}_j(t) =\hat{\mathcal{U}}^\dagger(t) \hat{\bm{p}}_j \hat{\mathcal{U}}(t)$ for simplicity. Under the assumptions in Eq. (\ref{eq:independent_and_identical_emitters}), the expression for the harmonic spectrum is given as (letting $t \rightarrow \infty$) \cite{Gorlach2020, Lange2024a}

\begin{align}
	S(\omega_k) &= \dfrac{\omega_k^3}{g_0^2 (2\pi)^2 c^3} \sum_\sigma \langle 	\hat{a}_{\bm{k}, \sigma}^\dagger 	\hat{a}_{\bm{k}, \sigma} \rangle \nonumber \\
	&= S_{\text{coh}}(\omega_k) + S_{\text{inc}}(\omega_k),  \label{eq:S_complete_exact}
\end{align}
where 

\begin{align}
	S_{\text{coh}}(\omega_k) &=  \dfrac{\omega_k^2}{(2\pi)^2 c^3}  N(N-1) \sum_\sigma \bigg \lvert \int_0^\infty dt' e^{-i \omega_k t'} \langle  \hat{{p}}_\sigma(t') \rangle \bigg \rvert^2, \label{eq:S_coh_exact} \\
	S_{\text{inc}}(\omega_k) &= \dfrac{\omega_k^2}{(2\pi)^2 c^3}  N \sum_\sigma  \int_0^\infty dt' \int_0^\infty dt'' \nonumber \\
	& \qquad \qquad \qquad \times e^{-i \omega_k (t'-t'')} \langle \hat{{p}}_\sigma(t')  \hat{{p}}_\sigma(t'') \rangle  \label{eq:S_inc_exact},
\end{align}
are the coherent part [proportional to $N(N-1)$] and incoherent part (proportional to $N$), respectively \cite{Sundaram1990, Stammer2025}, and where we denote $\hat{\bm{e}}_\sigma \cdot  \hat{\bm{p}}(t) =  \hat{{p}}_\sigma(t)$ as the projection of the momentum operator onto the polarization unit vector for notational convenience. We note in passing that typical HHG experiments consider systems with $N^2 \gg N$, which justifies only considering the coherent spectra [Eq. (\ref{eq:S_coh_exact})] in semiclassical HHG \cite{Sundaram1990}.

The degree of squeezing for a given mode $(\bm{k}, \sigma)$ is given by the minimum variance of the quadrature operator (in units of dB) \cite{Scully1997, Braunstein2005}
\begin{equation}
	\eta_{\bm{k}, \sigma} = - 10 \text{log}_{10} \bigg\{4 \underset{\theta_{\bm{k}, \sigma} \in [0, \pi)}{\text{min}}  [\Delta \hat{X}_{\bm{k}, \sigma}(\theta_{\bm{k}, \sigma})]^2 \bigg\}, \label{eq:squeezing_definition}
\end{equation}
where the quadrature operator is given as $\hat{X}_{\bm{k}, \sigma}(\theta_{\bm{k}, \sigma}) = (\hat{a}_{\bm{k}, \sigma} e^{-i \theta_{\bm{k}, \sigma}}+ \hat{a}_{\bm{k}, \sigma}^\dagger e^{i \theta_{\bm{k}, \sigma}})/2$. Using Eq. (\ref{eq:Heisenberg_solution_exact}), the quadrature variance is given as
\begin{align}
	& [\Delta \hat{X}_{\bm{k}, \sigma}(\theta_{\bm{k}, \sigma})]^2  = \dfrac{1}{4} + \dfrac{g_0^2 N}{2 \omega_k} \bigg\{ \nonumber \\
	&  + \int_0^t dt' \int_0^t dt'' e^{-i \omega_k (t'-t'')} \langle \Delta \hat{{p}}_\sigma(t')  \Delta \hat{{p}}_\sigma(t'') \rangle \nonumber \\
	& - \text{Re} \bigg[ e^{-2i (\omega_k t + \theta_{\bm{k}, \sigma})} \nonumber \\
	&\quad \times \int_{0}^t dt' \int_{0}^t dt'' e^{i \omega_k (t'+t'')} \langle \Delta \hat{{p}}_\sigma(t')   \Delta \hat{{p}}_\sigma(t'') \rangle \bigg] \bigg\}, \label{eq:quadrature_variance_exact}
\end{align}
where we have defined $\Delta \hat{{p}}_\sigma(t) = \hat{{p}}_\sigma(t)  - \langle \hat{{p}}_\sigma(t)  \rangle$ for notational convenience. We note that the quadrature variance scales linearly with the number of independent and phase-matched emitters, $N$. This is an important finding, motivating HHG as a method to generate squeezed states of light. This will be discussed further below. 

The photon statistics of a single mode of the emitted light can be quantified by the second-order intensity correlation function 
\begin{equation}
	g_{\bm{k}, \sigma}^{(2)}(0) = \frac{\langle \hat{a}_{\bm{k}, \sigma}^\dagger \hat{a}_{\bm{k}, \sigma}^\dagger \hat{a}_{\bm{k}, \sigma} \hat{a}_{\bm{k}, \sigma} \rangle}{ \langle \hat{a}_{\bm{k}, \sigma}^\dagger \hat{a}_{\bm{k}, \sigma} \rangle^2}. \label{eq:g2_general}
\end{equation}
An extension to a two-mode second-order correlation function is straightforward \cite{Walls2008}. If $g^{(2)}_{\bm{k}, \sigma}(0) = 1$, the light has Poissonian statistics, which is a characteristic of a coherent state. For $g^{(2)}_{\bm{k}, \sigma}(0) >1$, the light is super-Poissonian while $0 \leq g^{(2)}_{\bm{k}, \sigma}(0) < 1$ describes the exclusive quantum range of sub-Poissonian photon statistics. Using Eq. (\ref{eq:Heisenberg_solution_exact}), we can express the second-order correlation function as
\begin{equation}
	g_{\bm{k}, \sigma}^{(2)}(0) = \dfrac{N^4 C_{\bm{k}, \sigma}^{(0)} +f(N)}{N^4 C_{\bm{k}, \sigma}^{(0)} + h(N)}, \label{eq:g2_exact}
\end{equation}
where $C_{\bm{k}, \sigma}^{(0)}$ is the leading-order coefficient and where  $h(N)$ and $f(N)$ are different functions that both have leading-order terms of order $\mathcal{O}(N^3)$. The full expression for $g_{\bm{k}, \sigma}^{(2)}(0) $ is given in App. \ref{App:Photon_statistics}. As discussed in Ref. \cite{Stammer2025}, we also here find that for sufficiently large $N$, the emitted light emits Poissonian photon statistics, i.e., $g_{\bm{k}, \sigma}^{(2)}(0) \rightarrow 1$ for $N \rightarrow \infty$, independent of the nature of the generating emitter. However, one may design experiments where this high-$N$ limit is not reached, and hence, non-Poissonian statistics are obtainable. The value for $N$ depends on the nature of the generating medium and is discussed in Sec. \ref{Sec:Discussion}. Curiously, comparing Eqs. (\ref{eq:quadrature_variance_exact}) and (\ref{eq:g2_exact}), in the limit of a sufficiently large $N$, one may find that the light is highly squeezed while at the same time yielding classical Poissonian photon statistics. This classical limit in the photon statistics is similar to that of a Fock state, $\ket{n}$, which also has $g_{\bm{k}, \sigma}^{(2)}(0) \rightarrow 1$ for $n \rightarrow \infty$ while still yielding sub-Poissonian photon statistics. A calculation of the Mandel-Q parameter, $\mathcal{Q}_{\bm{k}, \sigma}$, is straight-forward through the relation $\mathcal{Q}_{\bm{k}, \sigma} = \langle \hat{a}_{\bm{k}, \sigma}^\dagger \hat{a}_{\bm{k}, \sigma} \rangle (g_{\bm{k}, \sigma}^{(2)}(0) -1 )$ \cite{Gerry2004}. We choose to consider $g_{\bm{k}, \sigma}^{(2)}(0)$ due to its popularity in experimental work \cite{Theidel2024, Lemieux2025, EvenTzur2025}. We refer the reader to App. \ref{App:Exp_electron_operators} for a derivation of how the $N$ scaling are calculated for the considered observables.

All the expressions for the observables of interest [Eqs. (\ref{eq:S_complete_exact}), (\ref{eq:quadrature_variance_exact}), and (\ref{eq:g2_exact})], require one to obtain $\hat{\bm{p}}(t) =\hat{\mathcal{U}}^\dagger(t)~ \hat{\bm{p}} ~ \hat{\mathcal{U}}(t)$, which can be obtained from the Heisenberg equation of motion for $\hat{\bm{p}}$. However, doing so will couple to the dynamics of the emitter back to the photonic operator [Eq. (\ref{eq:Heisenberg_solution_exact})], resulting in a hierarchy of higher and higher products of both electronic and photonic operators. While such products can be systematically factorized in an approximative manner \cite{KiraAndKoch2012, Gombkoto2016, Gombkoto2021}, we seek another route in this work and obtain a closed-form expression from a perturbative expansion of the time-evolution operator.
	
\subsection{Strong-field quantum optical perturbative Heisenberg dynamics (PHD)} \label{Sec:Perturbative_dynamics}

\subsubsection{Formalism}
In the strong-field quantum optical approach, the idea is to use (i) that the external coherent driving field is strong and equivalent to a classical field and (ii) the coupling to all other photonic modes is weak and can be treated in a perturbative manner. We first use the equivalence between a coherent state and a classical sinusoidal driving field to transfer the effect of the driving field from the initial state to the Hamiltonian via a displacement \cite{Cohen_Tannoudji1998_atomphoton}. This transformation, in turn, transforms the full Hamiltonian of the system [Eq. (\ref{eq:H_general_1})], which will then contain the purely semiclassical Hamiltonian, $\hat{H}_{sc}(t)$. By further going into the rotating frame of both the semiclassical Hamiltonian via $\hat{\mathcal{U}}_{sc}^\dagger(t)$ and the free-field Hamiltonian via $\mathcal{U}_F^\dagger(t)$, we obtain the state in the transformed frame $\ket{\tilde{\Psi}(t)} = \hat{\mathcal{U}}_{sc}^\dagger(t) \hat{D}^\dagger(\alpha) \mathcal{U}_F^\dagger(t) \ket{\Psi(t)}$, where all modes initially are in the vacuum state. This approach was adopted in early works on quantum optical HHG using the Schrödinger picture \cite{Gorlach2020} and has been widely used since. Consequently, the transformed expression for the expectation value of the operator $\hat{O}$ is given by (a full derivation is provided in App. \ref{App:expectation_value_expression})

\begin{equation}
	\langle \hat{O}(t) \rangle  = \bra{\tilde{\Psi}(0)} \mathcal{U}'^\dagger(t) \hat{O}'_S(t)\mathcal{U}'(t) \ket{\tilde{\Psi}(0)}, \label{eq:expectation_value_displaced}
\end{equation}
where 
\begin{equation}
	\mathcal{\hat{U}}'(t) = \hat{D}^\dagger(\alpha_{\bm{k}_L, \sigma_L})  \hat{\mathcal{U}}_{sc}^\dagger(t) \hat{\mathcal{U}}^\dagger_F(t) \mathcal{\hat{U}}(t) \hat{D}(\alpha_{\bm{k}_L, \sigma_L}) \label{eq:U_prime_exact}
\end{equation}
is the transformed time-evolution operator with $\hat{\mathcal{U}}(t)$ being the time-evolution operator in the original frame. In Eq. (\ref{eq:U_prime_exact}), $\hat{D}(\alpha_{\bm{k}_L, \sigma_L}) = \otimes_{\bm{k}, \sigma = \{(\bm{k}_L, \sigma_L)\}} \exp(\alpha_{\bm{k}, \sigma} \hat{a}_{\bm{k}, \sigma}^\dagger - \alpha_{\bm{k}, \sigma}^*\hat{a}_{\bm{k}, \sigma} )$ is the displacement operator of the modes present in the driving field and $\hat{\mathcal{U}}_{sc}(t)$ and $\hat{\mathcal{U}}^\dagger_F(t)$ are the time-evolution operators of the semiclassical Hamiltonian and the free-field Hamiltonian, respectively. The subscript, $S$, denotes the Schrödinger picture and the transformed operator is $ \hat{O}'_S(t) =  \hat{\mathcal{U}}_{sc}^\dagger(t) \hat{D}^\dagger(\alpha_{\bm{k}_L, \sigma_L})  \hat{\mathcal{U}}^\dagger_F(t) \hat{O}_S(t) \hat{\mathcal{U}}_F(t) \hat{D}(\alpha_{\bm{k}_L, \sigma_L})  \hat{\mathcal{U}}_{sc}(t)$. Finally, the initial state in the transformed frame is given by $\ket{\tilde{\Psi}(0)} = \hat{D}^\dagger(\alpha_{\bm{k}_L, \sigma_L}) \ket{\Psi(0)} = \otimes_{j=1}^N \ket{\phi_i^{(j)}} \otimes_{\bm{k}, \sigma} \ket{0_{\bm{k}, \sigma}}$, i.e., all photonic modes are initially in the vacuum state. 

Similarly to the case presented in Sec. \ref{Sec:exact_dynamics}, a Heisenberg equation of motion exists for the photonic operator $\hat{a}'_{\bm{k}, \sigma}(t)$. The solution to such a Heisenberg equation of motion is equivalent to Eq. (\ref{eq:Heisenberg_solution_exact}), given as

\begin{equation}
	\hat{a}_{\bm{k}, \sigma}'(t) = \hat{a}'_{\bm{k}, \sigma}(0) e^{-i \omega_k t} - i \dfrac{g_0}{\sqrt{\omega_k}} \int_{0}^{t} dt' \hat{\bm{e}}_\sigma \cdot \hat{\bm{Q}}'(t') e^{-i\omega_k(t-t')}, \label{eq:Heisenberg_transformed_exact}
\end{equation}	
where 
\begin{equation}
	\hat{\bm{Q}}'(t) = \hat{\mathcal{U}}'^\dagger(t) 	\hat{\bm{Q}}_{sc}(t) \hat{\mathcal{U}}'(t), \label{eq:Q_H_def}
\end{equation}
describes the exact dynamics of the driven emitter, which, due to the applied transformations, is now expressed as the semiclassically driven emitter, $\hat{\bm{Q}}_{sc}(t)$, coupled to the photonic field via the transformed time-evolution operator $\hat{\mathcal{U}}'(t)$. The expression for the semiclassically driven emitter is given as 
\begin{align}
	\hat{\bm{Q}}_{sc}(t) &= \hat{\mathcal{U}}_{sc}^\dagger(t) \sum_{j=1}^N [\hat{\bm{p}}_j + \bm{A}_{cl}(t)] \hat{\mathcal{U}}_{sc}(t) \nonumber \\
	&= \sum_{j=1}^N [\hat{\bm{p}}_{j,sc}(t)+ \bm{A}_{cl}(t)], \label{eq:Q_definition}
\end{align}
and is the sum of single-emitter momentum operators, where we for convenience have defined the semiclassically time-evolved momentum operator for emitter $j$ as 
\begin{equation}
	\hat{\bm{p}}_{j,sc}(t) =  \hat{\mathcal{U}}_{sc}^\dagger(t)  \hat{\bm{p}}_j  \hat{\mathcal{U}}_{sc}(t), \label{eq:p_sc}
\end{equation}
and where $\bm{A}_{cl}(t) = \bra{\alpha_{\bm{k}_L, \sigma_L}} \hat{\bm{A}}(t) \ket{\alpha_{\bm{k}_L, \sigma_L}}$ is the classical driving field with $\hat{\bm{A}}(t)$ given in Eq. (\ref{eq:A_operator_2}) below. Equation (\ref{eq:Heisenberg_transformed_exact}) is exact in the transformed frame and equivalent to Eq. (\ref{eq:Heisenberg_solution_exact}). Unfortunately, Eq. (\ref{eq:Heisenberg_transformed_exact}) still requires the full time-evolution operator for the combined light-matter system, as seen from Eqs. (\ref{eq:U_prime_exact}) and (\ref{eq:Q_H_def}), and we face the same challenges as discussed in Sec. \ref{Sec:exact_dynamics}. To proceed, we seek to obtain an approximative but closed-form expression of Eq. (\ref{eq:Heisenberg_transformed_exact}).

From Eq. (\ref{eq:U_prime_exact}), it can be shown (see also App. \ref{App:expectation_value_expression}), that the time-evolution operator in Eq. (\ref{eq:expectation_value_displaced}) satisfies the time-dependent Schrödinger equation (TDSE)
\begin{equation}
	\hat{\mathcal{U}}'(t) = \mathbb{1} - i \int_0^t dt' \hat{V}(t') 	\hat{\mathcal{U}}'(t'), \label{eq:U_prime_int_eq}
\end{equation}
with the interaction term given by
\begin{equation}
	\hat{V}(t) = \hat{\bm{A}}(t) \cdot \hat{\bm{Q}}_{sc}(t), \label{eq:V_definition}
\end{equation}
where
\begin{equation}
	\hat{\vec{A}}(t) = \sum_{\bm{k}, \sigma} \dfrac{g_0}{\sqrt{\omega_k}} \big[\hat{a}_{\bm{k}, \sigma} (0) e^{-i \omega_k t} + \hat{a}_{\bm{k}, \sigma}^\dagger (0) e^{i \omega_k t} \big] \hat{\bm{e}}_\sigma, \label{eq:A_operator_2}
\end{equation}
is the vector potential operator, which is now time dependent due to the transformation to the frame of the free-field Hamiltonian.

We seek to expand Eq. (\ref{eq:U_prime_int_eq}) in a perturbative manner in orders of the light-matter coupling $g_0$. A comment is worth making here. In the case presented in Sec. \ref{Sec:exact_dynamics}, such a perturbative expansion is not possible, as $\abs{\alpha_L g_0} \gtrsim 1$ since $\lvert \alpha_L \rvert \gg 1$ due to the intense driving field. Now, we have, via transformations, put the effect of the driving field in the Hamiltonian such that the semiclassical dynamics are accounted for via the transformation using $\hat{\mathcal{U}}_{sc}(t)$. Consequently, all photonic modes now refer to the vacuum state, which in turn allows the perturbative expansion to be in orders of the coupling to the generated field, which is typically small. We hence expand Eq. (\ref{eq:U_prime_int_eq}) to second order in $N g_0 \lvert \tilde{\bm{p}}(\omega) \rvert$, where $\lvert \tilde{\bm{p}}(\omega) \rvert < 5$ a.u. is the typical dipole response from a single emitter in the atomic and strongly correlated systems considered below. As also discussed below, we find that the inequality
\begin{equation}
	N g_0 \lvert \tilde{\bm{p}}(\omega) \rvert \ll 1, \label{eq:perturbative_inequality}
\end{equation}
 is the relevant regime for the number of emitters, $N$, present in HHG experiments, i.e., high-order terms in this perturbative expansion do not contribute to measurable observables in typical HHG experiments. 

We perform the perturbative expansion by considering the time-evolution operator $\hat{\mathcal{U}}'(t)$ whose equation is given in Eq. (\ref{eq:U_prime_int_eq}). We now expand Eq. (\ref{eq:U_prime_int_eq}) to second order in the interaction, $\hat{V}(t)$, and obtain
\begin{equation}
	\hat{\mathcal{U}}'(t) \simeq \mathbb{1} -i \int_{0}^t dt' \hat{V}(t') - \int_0^t dt' \int_0^{t'} dt'' \hat{V}(t') \hat{V}(t''). \label{eq:U_prime_expansion_second_order}
\end{equation}
From Eq. (\ref{eq:V_definition}), we note that each $\hat{V}(t)$ carries a factor of the light-matter coupling $g_0$ as well as a sum over all independent emitters, effectively carrying an additional factor of $N  \lvert \tilde{\bm{p}}(\omega) \rvert$.
By inserting Eq. (\ref{eq:U_prime_expansion_second_order}) into Eq. (\ref{eq:Q_H_def}), we obtain
\begin{align}
	\hat{\bm{Q}}'(t)& = \hat{\bm{Q}}_{sc}(t) + \hat{\bm{Q}}_q(t), \label{eq:Q_prime_both_contributions} 
\end{align}
with
\begin{align}
	\hat{\bm{Q}}_q(t) &\simeq 	\hat{\bm{Q}}_q^{(1)}(t) +  \hat{\bm{Q}}_q^{(2)}(t), \label{eq:Q_q_terms}
\end{align}
being the quantum correction to the semiclassical dynamics, $\hat{\bm{Q}}_{sc}(t)$, which occurs due to the coupling to the quantized electromagnetic field. To second order in the light-matter coupling, $\hat{V}(t)$, the quantum corrections to the semiclassical dynamics in Eq. (\ref{eq:Q_q_terms}) are given by
\begin{align}
	\hat{\bm{Q}}_q^{(1)}(t)  &= i \int_0^t dt' [\hat{V}(t'), \hat{\bm{Q}}_{sc}(t)], \nonumber \\
	\hat{\bm{Q}}_q^{(2)}(t)  &=	 \int_0^t dt' \int_0^t dt'' \hat{V}(t') \hat{\bm{Q}}_{sc}(t) \hat{V}(t'') \nonumber \\
	& \quad - \int_0^t \! \! dt' \! \! \int_0^{t'} \! \! dt'' \left[ \hat{\bm{Q}}_{sc}(t) \hat{V}(t') \hat{V}(t'')  + \text{h.c.}\right], \label{eq:Q_H_expansion}
\end{align}
where the order in $g_0 N  \lvert \tilde{\bm{p}}(\omega) \rvert $ is indicated by their superscripts and the definition of $ \hat{\bm{Q}}_{sc}(t)$ is given in Eq. (\ref{eq:Q_definition}). 

\begin{figure}
	\centering
	\includegraphics[width=1\linewidth]{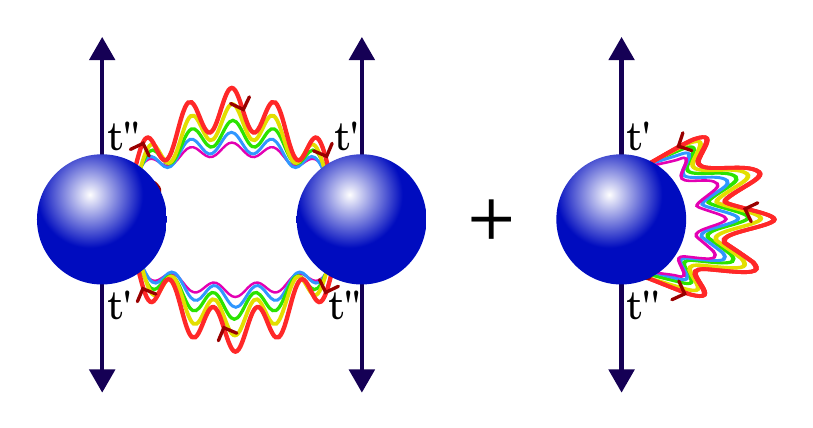}
	\caption{Quantum corrections to semiclassical emitter dynamics. Calculating $\langle \hat{\bm{Q}}'(t) \rangle$ from Eqs. (\ref{eq:Q_prime_both_contributions})-(\ref{eq:Q_H_expansion}), the strong-field quantum optical PHD expressions for the emitter (blue circles) dynamics show that a correction term to the semiclassical dynamics (blue arrows) is causing photon exchanges of all modes (colored curves) between both different and the same emitter for times $t'' < t'$, due to the coupling to the photonic field. In this sense, the emitters experience beyond-semiclassical dynamics, which in turn affects the nature of the generated light in HHG.}
	\label{fig:dipolephotoninteraction}
\end{figure}

 Note that the prime of $\hat{\bm{Q}}'(t)$ is not found on the right-hand side of Eq. (\ref{eq:Q_prime_both_contributions}). We choose this notation in order to emphasize that the exact dynamics in the transformed frame, denoted by the prime, is described by the semiclassical dynamics with additional quantum-induced corrections. 

Equations (\ref{eq:Q_prime_both_contributions})-(\ref{eq:Q_H_expansion}) capture the description of the dynamics of the emitter using this strong-field quantum optical perturbative Heisenberg dynamics (PHD) and form one of the central sets of equations of this work. 

By calculating the expectation value of the emitter dynamics in Eq. (\ref{eq:Q_prime_both_contributions}), we find that the semiclassical emitter dynamics are corrected due to the coupling to the quantized field. This is shown in Fig. \ref{fig:dipolephotoninteraction}, where we illustrate how an emitter couples both to itself and to other emitters via photonic exchanges for $t'' < t'$, even though there is no explicit interaction between the emitters in the Hamiltonian [Eq. (\ref{eq:H_general_1})]. This photon exchange both affects the mean and variance of the emitter dynamics. The PHD description thus includes beyond-semiclassical emitter dynamics, which have not previously been considered in strong-field quantum optics. 

\subsubsection{Backaction and photonic observables}

Returning to the photonic operator, we insert Eq. (\ref{eq:Q_H_expansion}) into Eq. (\ref{eq:Heisenberg_transformed_exact}) and obtain
\begin{align}
	\hat{a}_{\bm{k}, \sigma}'(t) = & ~\hat{a}'_{\bm{k}, \sigma}(0) e^{-i \omega_k t} \nonumber \\
	&-i \dfrac{g_0}{\sqrt{\omega_k}}\int_{0}^{t}\!\! dt' \hat{\bm{e}}_\sigma \cdot \big[\hat{\bm{Q}}_{sc}(t') + \hat{\bm{Q}}_q(t') \big] e^{-i\omega_k(t-t')}. 	\label{eq:Heisenberg_perturbative}
\end{align}
Equation (\ref{eq:Heisenberg_perturbative}) is the solution for the photonic operator using the PHD description and is the central equation describing the photon dynamics. This equation is a closed-form expression for the photonic operator and includes the backaction of the emitter onto the driving field due to both semiclassical dynamics and its corrections due to the coupling to the quantized electromagnetic field. Only using $\hat{\bm{Q}}_{sc}(t)$ instead of $\hat{\bm{Q}}'(t)$ in Eq. (\ref{eq:Heisenberg_transformed_exact}), i.e., replacing the exact dynamics of the driven emitters with the semiclassical dynamics, would be sufficient to account for the generation of nonclassical light. However, as will be clear below, it would not capture higher-order terms necessary to consistently calculate the nonclassical observables of interest to order $\mathcal{O}(g_0^3)$. The effect of the quantum corrections to the semiclassical dynamics of the emitter on the nonclassicality of the emitted light is investigated in Sec. \ref{Sec:QuantumCorrectionsToSemiclassical}.

In the following we neglect the fundamental mode in the consideration of observables. This case requires special handling due to the fact that $\bm{A}_{cl}(t)$ enters in the expression for $\hat{\bm{Q}}_{sc}(t)$ as seen in Eq. (\ref{eq:Q_definition}). It is straightforward to include this case, but it is omitted for simplicity in the equations.

Using the PHD expression for the photonic field in Eq. (\ref{eq:Heisenberg_perturbative}), we calculate the observables of interest. Using the assumption of identical and independent emitters given in Eq. (\ref{eq:independent_and_identical_emitters}), the harmonic spectrum is given by

\begin{align}
	S(\omega_k) = S_{\text{coh}}(\omega_k) + S_{\text{inc}}(\omega_k) + \mathcal{O}(g_0^2 N^3), \label{eq:spectrum_perturbative}
\end{align}
where
\begin{align}
	S_{\text{coh}}(\omega_k) &= \dfrac{\omega_k^2}{(2\pi)^2 c^3}  N(N-1) \sum_\sigma \bigg \lvert \int_0^\infty dt' e^{-i \omega_k t'} \langle  \hat{{p}}_{sc,\sigma}(t') \rangle \bigg \rvert^2,\label{eq:spectrum_perturbative_coherent} \\
	S_{\text{inc}}(\omega_k) &= \dfrac{\omega_k^2}{(2\pi)^2 c^3}  N \sum_\sigma  \int_0^\infty dt' \int_0^\infty dt'' \nonumber \\
	& \qquad \qquad \qquad e^{-i \omega_k (t'-t'')} \langle \hat{{p}}_{sc, \sigma}(t')  \hat{{p}}_{sc,\sigma}(t'')  \rangle, \label{eq:spectrum_perturbative_incoherent}
\end{align}
and where $\hat{{p}}_{sc, \sigma}(t) = \hat{\bm{e}}_{\sigma} \cdot \hat{\bm{p}}_{sc}(t)$ is given in Eq. (\ref{eq:p_sc})

Interestingly, to leading order in $g_0 N  \lvert \tilde{\bm{p}}(\omega) \rvert$, the PHD-predicted spectrum in Eq.~(\ref{eq:spectrum_perturbative}) coincides with the exact expression in Eq.~(\ref{eq:S_complete_exact}) if the exact dynamics of the emitter is replaced by their semiclassical form. This means that the quantum corrections to the emitter dynamics are not seen in the harmonic spectrum and only the semiclassical dynamics are relevant for this observable.

As shown in App. \ref{App:higher_order_corrections}, the next correction to the harmonic spectrum, of order $\mathcal{O}(g_0^2 N^3)$, violates the symmetry-based selection rules obeyed by the coherent contribution in Eq.(\ref{eq:spectrum_perturbative_coherent}). Since HHG experiments measure only odd harmonics from, e.g., inversion-symmetric targets \cite{Neufeld2019}, this shows that the coherent contribution to the spectrum is dominant, which puts an upper and lower bound on the value of $N$. The coherent contribution [Eq. (\ref{eq:spectrum_perturbative_coherent})] must dominate the incoherent contribution [Eq. (\ref{eq:spectrum_perturbative_incoherent})] meaning that it must hold that $N^2 \gg N$. On the other hand, the coherent contribution to the spectrum must also dominate the correction term, and it must hence be the case that $N^2 \gg N^3 g_0^2$ consistent with $N g_0 \lvert \tilde{\bm{p}}(\omega) \rvert \ll 1$ discussed above. If instead $N^2 \ll  g_0^2 N^3$, the noncompliant $ g_0^2 N^3$ term would dominate, and HHG only at odd harmonics from, e.g., inversion-symmetric systems, would not be observed. In other words, the fact that experiments only observe odd harmonics in the symmetric pulse limit and for inversion-symmetric samples proves that the coherent part of the spectrum is the dominating term, which puts both an upper and lower boundary on $N$ for a given $g_0$. Consequently, this shows that the PHD is valid in the regime relevant for typical HHG experiments. A discussion of the value of $g_0$ and $N$ is given in Sec. \ref{Sec:Results}.	

Turning to the PHD expression for the degree of squeezing, we find that, to leading order in $g_0N  \lvert \tilde{\bm{p}}(\omega) \rvert$, the variance of the quadrature operator is given as
\begin{align}
	& [\Delta \hat{X}_{\bm{k}, \sigma}(\theta_{\bm{k}, \sigma})]^2  = \dfrac{1}{4} + \dfrac{g_0^2 N}{2 \omega_k} \bigg\{ \nonumber \\
	&\int_0^t dt' \int_0^t dt'' e^{-i \omega_k (t'-t'')} \langle \Delta \hat{p}_{sc, \sigma}(t') \Delta \hat{p}_{sc, \sigma}(t'') \rangle \nonumber \\
	& 
	- \text{Re} \bigg[  \int_0^t dt' \int_0^{t} dt'' e^{-i \omega_k (2t -t' -t'') - 2i \theta_{\bm{k}, \sigma}} \nonumber \\
	& \qquad \qquad \qquad \qquad \times \langle \Delta \hat{p}_{sc, \sigma}(t') \Delta \hat{p}_{sc, \sigma}(t'') \rangle  \nonumber\\
	& -	\int_0^t dt' \int_0^{t'} dt'' e^{-i \omega_k (2t -t' -t'') - 2i \theta_{\bm{k}, \sigma}} \nonumber \\
	& \qquad \qquad \qquad \qquad \times  \langle [\Delta \hat{p}_{sc, \sigma}(t''), \Delta \hat{p}_{sc, \sigma}(t')] \rangle
	\bigg] \bigg\}. \label{eq:quadrature_variance_peturbative}
\end{align}
where we have defined $\Delta \hat{{p}}_{sc, \sigma}(t) = \hat{{p}}_{sc, \sigma}(t)  - \langle \hat{{p}}_{sc, \sigma}(t) \rangle$ for notational convenience. We note that the first two terms in Eq. (\ref{eq:quadrature_variance_peturbative}) are due to the semiclassical dynamics of the emitter. In contrast, the upper limit of the last integral in Eq. (\ref{eq:quadrature_variance_peturbative}) is $t'$, and this term is due to the first-order quantum correction of the semiclassical dynamics given in Eq. (\ref{eq:Q_H_expansion}). Importantly, this shows that the degree of squeezing is due to both the time correlations in the semiclassical dynamics of the emitter, but also due to the fluctuations of the semiclassical dynamics induced by the coupling to the quantized field. Hence, to obtain an expression for the degree of squeezing consistent in orders of the coupling, one needs to include the quantum-induced fluctuations on the semiclassical dynamics, i.e., $\hat{\bm{Q}}_q(t)$. We note that Eq. (\ref{eq:quadrature_variance_peturbative}) can be simplified via time-ordering considerations, but we express the quadrature variance as in Eq. (\ref{eq:quadrature_variance_peturbative}) to clearly identify contributions due to the beyond-semiclassical dynamics of the emitter.

Within the PHD, the second-order correlation function is given by	
\begin{equation}
	g_{\bm{k}, \sigma}^{(2)}(0) = \dfrac{N^4 D_{\bm{k}, \sigma}^{(0)} +f_\text{PHD}(N)}{N^4 D_{\bm{k}, \sigma}^{(0)} + h_\text{PHD}(N)} \label{eq:g2_perturbative}
\end{equation}
where $D_{\bm{k}, \sigma}^{(0)}$ is the leading-order coefficient and the full expression is given in App. \ref{App:Photon_statistics}. The functions $f_\text{PHD}(N)$ and $h_\text{PHD}(N)$ both contain terms on order $\mathcal{O}(N^3)$ while being different functions. We emphasize that Eq. (\ref{eq:g2_perturbative}) is independent of the coupling constant $g_0$ when expanding both numerator and denominator consistently to order $\mathcal{O}(g_0^4)$.

A number of general comments about the PHD expressions for the spectrum [Eq. (\ref{eq:spectrum_perturbative})], quadrature variance related to the degree of squeezing [Eq. (\ref{eq:quadrature_variance_peturbative})] and photon statistics [Eq. (\ref{eq:g2_perturbative})] are in order. Firstly, we note that the expressions in the PHD all show the same scaling with the number of independent emitters, $N$, as the corresponding exact expressions in Sec. \ref{Sec:exact_dynamics}. Notably, we find that the degree of squeezing is increasing with $N$. Secondly, we note that the PHD, to leading order in $g_0N  \lvert \tilde{\bm{p}}(\omega) \rvert$, shows that the different frequency components do not couple, i.e., the expressions for the spectrum, squeezing and $g_{\bm{k}, \sigma}^{(2)}(0)$ calculated for the mode $(\bm{k}, \sigma)$ are independent of all other modes $(\bm{k}', \sigma') \neq (\bm{k}, \sigma)$. Finally, we refer to App. \ref{App:Exp_electron_operators} where it is shown how the different electronic expectation values are calculated, including their respective scaling with $N$.

\section{Results} \label{Sec:Results}
We illustrate the generality of the strong-field quantum optical PHD by considering its use for modeling of the nonclassical response from two different types of emitter systems, namely an ensemble of atoms and a strongly correlated material described by the Fermi-Hubbard Hamiltonian. We consider linearly polarized light and capture the essential parts of the dynamics along the polarization of the external coherent driver by one-dimensional models of the electronic systems. We calculate the harmonic spectrum [Eq. (\ref{eq:spectrum_perturbative})], the degree of squeezing from the variance of the quadrature operator [Eq. (\ref{eq:quadrature_variance_peturbative})] and the second-order correlation function [Eq. (\ref{eq:g2_perturbative})]. We note that, to leading order, both the harmonic spectrum and the second-order correlation function are independent of the value of $g_0$, while the variance in the quadrature operator depends on $\propto g_0^2 N  = 2\pi N/V$, i.e., it depends on the density of the emitters considered. We choose to use a quantization volume of $V = (10 \lambda_L)^3$ and emphasize that the corresponding numerical value of $g_0$ does not qualitatively change the results as long as Eq. (\ref{eq:perturbative_inequality}) is obeyed. 

\begin{figure}
	\centering
	\includegraphics[width=1\linewidth]{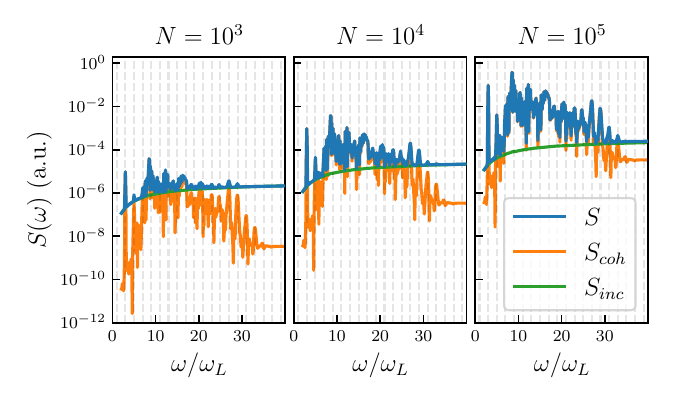}
	\caption{HHG spectra calculated using Eq. (\ref{eq:spectrum_perturbative}) for an atomic ensemble of $N = 10^3, 10^4$, and $10^5$ atoms. The total spectrum (blue) is dominated by the incoherent (green) part for $N=10^3$, while the coherent spectrum (orange) is dominating for $N \geq 10^4$. Note that the fundamental mode is not shown as discussed in the main text. See Sec. \ref{Sec:Res_AtomicEnsemble} for system parameters.}
	\label{fig:atomspektre}
\end{figure}

\subsection{Procedure for calculating photonic observables} \label{Sec:Res_procedure}
Within the strong-field quantum optical PHD framework, photonic observables such as the harmonic spectrum [Eq.~(\ref{eq:spectrum_perturbative})], the quadrature variance used to quantify squeezing [Eq.~(\ref{eq:quadrature_variance_peturbative})], and photon statistics [Eq.~(\ref{eq:g2_perturbative})] can be evaluated. Common to these is the fact that they rely on expectation values derived from semiclassical theory, enabling the use of well-established semiclassical TDSE solvers. In semiclassical theory, the evaluation of single-time expectation values such as
\begin{equation}
	\langle \hat{p}_{sc,\sigma}(t) \rangle
	= \bra{\phi_i(t)} \hat{p}_{\sigma} \ket{\phi_i(t)},
	\quad
	\ket{\phi_i(t)} = \hat{\mathcal{U}}_{sc}(t) \ket{\phi_i(0)},
\end{equation}
is straightforward, where $i$ denotes the initial state of the emitter system, $\hat{\mathcal{U}}_{sc}(t)$ denotes the semiclassical time-evolution operator, and $\hat{p}_{\sigma}$ is the momentum operator projected onto the axis of the polarization $\sigma$. In contrast, the evaluation of two-time correlation functions, \(\langle \Delta \hat{p}_{sc}(t')\,\Delta \hat{p}_{sc}(t'') \rangle\), is more involved.

To compute these correlations, we insert the resolution of the identity,
\(\mathbb{1} = \sum_m |\phi_m\rangle\langle \phi_m|\),
where \(|\phi_m\rangle\) are field-free eigenstates of the emitter. This yields (see also App.~\ref{App:Exp_electron_operators})
\begin{align}
	&\langle \Delta \hat{{p}}_{sc, \sigma}(t')\Delta \hat{{p}}_{sc, \sigma}(t'') \rangle \nonumber \\
	&= \langle \hat{{p}}_{sc, \sigma}(t') \hat{{p}}_{sc, \sigma}(t'') \rangle - \langle \hat{{p}}_{sc, \sigma}(t') \rangle \langle \hat{{p}}_{sc, \sigma}(t'')\rangle \nonumber \\
	&= \sum_m {p}^{(\sigma)}_{i,m}(t') {p}^{(\sigma)}_{m,i}(t'') - {p}^{(\sigma)}_{i,i}(t'){p}^{(\sigma)}_{i,i}(t'') \nonumber \\
	&= \sum_{m\neq i} {p}^{(\sigma)}_{i,m}(t'){p}^{(\sigma)}_{m,i}(t''), \label{eq:correlations_to_cross_dipoles}
\end{align}
where $\langle \hat{p}_{sc,\sigma}(t) \rangle = p^{(\sigma)}_{i,i}(t)$ is the expectation value of the single dipole along the polarization direction \(\hat{\bm{e}}_{\sigma}\). The transition dipole matrix elements are given explicitly by
\begin{equation}
	p^{(\sigma)}_{m,n}(t)
	= \langle \phi_m(0) |
	\hat{\mathcal{U}}_{sc}^\dagger(t)\,\hat{{p}}_{\sigma}\,\hat{\mathcal{U}}_{sc}(t)
	| \phi_n(0) \rangle,
	\label{eq:p_mn}
\end{equation}
with $\hat{{p}}_{\sigma} = \hat{\bm{p}} \cdot \hat{\bm{e}}_{\sigma}$. Equation (\ref{eq:p_mn}) shows that the evaluation of the time-correlation functions required for the nonclassical response in HHG requires all the field-free eigenstates of the emitter and the semiclassical propagation of each of them. This requirement puts a constraint on the size of the emitter system due to the growing size of the Hilbert space of the emitter. In practice, this limitation can be mitigated by identifying and retaining only those transition dipole matrix elements that significantly contribute to the observable and frequency range of interest.

In summary, within the strong-field quantum-optical PHD approach, the transition dipole matrix elements are first computed using Eq.~(\ref{eq:p_mn}), which requires semiclassical TDSE propagation of the emitter eigenstates. Through Eq.~(\ref{eq:correlations_to_cross_dipoles}), these matrix elements determine the relevant time-correlation functions. The resulting quantities are then substituted into the expressions for the desired observables [Eqs.~(\ref{eq:spectrum_perturbative}), (\ref{eq:quadrature_variance_peturbative}), and (\ref{eq:g2_perturbative})], and the remaining integrals are evaluated by standard numerical integration.

\subsection{Atomic ensemble}\label{Sec:Res_AtomicEnsemble}

To capture the characteristics of the light generated from an atomic ensemble, we calculate the HHG response from a one-dimensional atomic model. To model this, we use the atomic potential $U(x)=-1/\sqrt{x^2+\epsilon^2}$ with $\epsilon=0.816$ a.u. to match the ionization potential of Ne ($I_p = 0.7926$ a.u.). A similar model was used in Ref. \cite{Gorlach2023}. We drive the system with a coherent laser pulse $A_{cl}(t) = (F_0/\omega_L) \sin^2[\omega_L t / (2 N_c)] \sin(\omega_L t +\pi /2)$ with $F_0=0.053$ a.u. corresponding to a peak intensity of $10^{14}$ W$/\text{cm}^2$ and angular frequency of $\omega_L=0.057$ a.u. corresponding to a wavelength of $800$ nm. The pulse duration is $N_c=20$ cycles. The semiclassical electron dynamics are obtained using a Crank-Nicolson routine \cite{Bauer2017computational} where the $x$-axis is discretized from $-3 \times 10^3$ a.u. to $3 \times 10^3$ a.u. with $1.5\times 10^5$ points. The used time step is $dt=0.02$ a.u., and absorbing boundary conditions are implemented by the multiplication of a masking function after every propagation step. For the results presented in this system, a value of $g_0 = 4 \times 10^{-8}$ a.u. is used, similar to the one used in \cite{Gorlach2020, Lewenstein2021, Lange2024a, Lange2025a, Lange2025b}. We note that the large box for the electron dynamics is required to calculate the time correlations of the induced dipole, as it requires the time propagation of the delocalized excited states of the field-free system, as explained in Sec. \ref{Sec:Res_procedure} and App. \ref{App:Exp_electron_operators}. Due to the fact that excited states are more likely to ionize, a larger box is required to contain the population of the free electron. In this model, we find convergence when including the $250$ lowest field-free eigenstates of the atomic system, which includes all bound states and the lowest-lying continuum states. 

\begin{figure}
	\centering
	\includegraphics[width=1\linewidth]{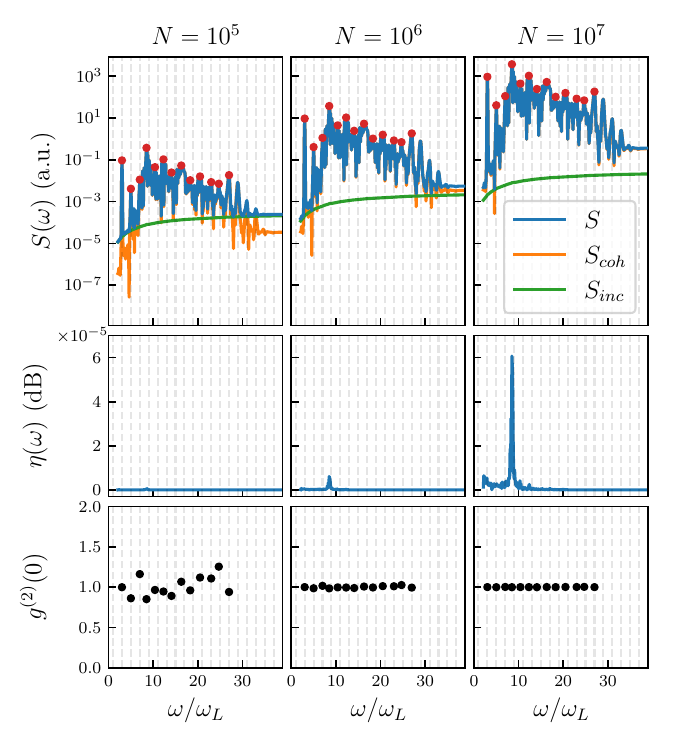}
	\caption{Response from an atomic ensemble of $N=10^5, 10^6$ and $ 10^7$ atoms with parameters given in the main text. The HHG spectra (top row), calculated using Eq. (\ref{eq:spectrum_perturbative}) show the contributions from both the incoherent [Eq. (\ref{eq:spectrum_perturbative_incoherent})] and coherent [Eq. (\ref{eq:spectrum_perturbative_coherent})] part. The values of $N$ are chosen to be in a regime where the coherent part is the dominant contribution. The squeezing (middle row) is calculated using Eq. (\ref{eq:quadrature_variance_peturbative}). We see that the degree of squeezing increases with $N$ as predicted. We find that a resonance in the atomic energy levels gives rise to a large degree of squeezing at close to the $9$'th harmonic. The second-order correlation function (bottom row) is calculated using Eq. (\ref{eq:g2_perturbative}) at frequencies indicated by red dots in the top row. Both sub- and super-Poissonian photon statistics are found for lower values of $N$, but tend towards classical Poissonian statistics for increasing values of $N$. Note that the fundamental mode is not shown as discussed in the main text. See Sec. \ref{Sec:Res_AtomicEnsemble} for system parameters.}
	\label{fig:atomfullfigure}
\end{figure}

In Fig. \ref{fig:atomspektre}, we show three different HHG spectra for $N = 10^3, 10^4,$ and $10^5$ independent atomic emitters. We see how the incoherent contribution to the spectrum [Eq. (\ref{eq:spectrum_perturbative_incoherent})] dominates the total spectrum for $N=10^3$ (especially at $\omega/\omega_L \geq 15$), while the coherent contribution [Eq. (\ref{eq:spectrum_perturbative_coherent})] dominates for $N \geq 10^4$ at most frequencies. Using experimental results as a reference where only the coherent spectrum seems to be observed, we limit ourselves in the following to $N > 10^4$ while still satisfying Eq. (\ref{eq:perturbative_inequality}) as discussed in Sec. \ref{Sec:Perturbative_dynamics}. With these experimentally realistic constraints for $N$, we show the harmonic spectra as well as the degree of squeezing [Eq. (\ref{eq:quadrature_variance_peturbative})] and second-order correlation function [Eq. (\ref{eq:g2_perturbative})] for $N = 10^5, 10^6$, and $10^7$ in Fig. \ref{fig:atomfullfigure}. Here, the top row shows the HHG spectra along with the different contributions, the middle row shows the degree of squeezing, and the bottom row shows the second-order correlation function. Looking at the spectrum, we note that the signal is predominantly on the odd harmonics and that the signal is stronger for more emitters. This behavior is expected for the coherent contribution to the spectrum \cite{Sundaram1990}. In the second row of Fig. \ref{fig:atomfullfigure}, we note how the degree of squeezing increases with the number of emitters as predicted by Eq. (\ref{eq:quadrature_variance_peturbative}), albeit it remains fairly small. Interestingly, we find a strong signal in the squeezing around the $9$'th harmonic, which we attribute to a $9$-photon resonance in the atomic states. We find that a clear squeezing is always found at the lowest atomic resonance, independent of the laser frequency. Crucially, such a resonance is not taken into account in an SFA treatment of the semiclassical dynamics, showing that approaches successful in semiclassical theory are not directly applicable in a quantum-optical treatment.

Looking at the second-order correlation function [Eq. (\ref{eq:g2_perturbative})] in the bottom row of Fig. \ref{fig:atomfullfigure}, we see how $g^{(2)}(0) \approx 1$ for $N=10^6$ and $N = 10^7$, while for $N = 10^5$ both super- and sub-Poissonian photon statistics is predicted. The fact that some harmonics yield super-Poissonian statistics while others have sub-Poissonian statistics, in general, depends on the nature of the generating medium, and it is beyond the scope of the current work to discuss these differences in detail. 

\subsection{Strongly correlated material} \label{Sec:Res_StronglyCorrelatedMaterial}

\begin{figure}
	\centering
	\includegraphics[width=1\linewidth]{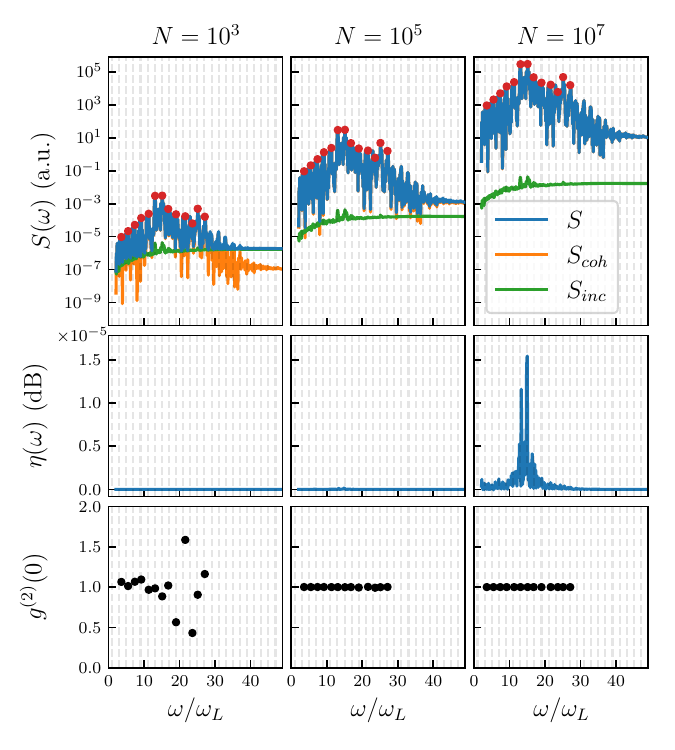}
	\caption{Response from $N=10^3, 10^5$, and $10^7$ (left to right) independent systems of the Fermi-Hubbard model with parameters given in the main text. The HHG spectra (top row) calculated using Eq. (\ref{eq:spectrum_perturbative}) show how the coherent contribution to the spectrum dominates for all considered values of $N$. In particular, we see a clear signal around the exciton energy. The degree of squeezing (middle row) is calculated via the quadrature variance in Eq. (\ref{eq:quadrature_variance_peturbative}), and we see a clear signal at the exciton energy with a stronger signal for increasing values of $N$. The second-order correlation function (bottom row), calculated using Eq. (\ref{eq:g2_perturbative}), shows that only for sufficiently small values of $N \approx 10^3$, a deviation from Poissonian statistics can be expected. For larger values of $N \geq 10^5$, the photon statistics approach a Poissonian distribution. Note that the fundamental mode is not shown as discussed in the main text. See Sec. \ref{Sec:Res_StronglyCorrelatedMaterial} for system parameters.}
	\label{fig:fhfullfigure}
\end{figure}
To showcase the generality of the PHD, we similarly calculate the observables of interest for a strongly correlated material modeled by the extended $U$-$V$ Fermi-Hubbard model. We consider the Hamiltonian
\begin{equation}
	\hat{H}_{sc}(t) = -t_0 \sum_{i, \mu} \bigg( e^{i a A_{cl}(t)} \hat{c}_{i, \mu}^\dagger \hat{c}_{i+1, \mu} + \text{h.c.} \bigg) + \hat{H}_{e-e}, \label{eq:H_Hubbard_full}
\end{equation}
where the electron-electron interaction is given by
\begin{align}
	\hat{H}_{e-e} = U \sum_i \hat{n}_{i, \uparrow} \hat{n}_{i, \downarrow} + V  \sum_{i} \hat{n}_{i} \hat{n}_{i+1}, \label{eq:H_ee}
\end{align}
where $\mu = \uparrow, \downarrow$ is the spin orientations, $\hat{n}_{i, \mu} =  \hat{c}_{i, \mu}^\dagger \hat{c}_{i \mu}$ is the electron counting operator on site $i$, and $\hat{n}_i = \hat{n}_{i, \uparrow} + \hat{n}_{i, \downarrow}$. In Eq. (\ref{eq:H_Hubbard_full}), $t_0$ is a nearest-neighbor hopping parameter, $a$ is the lattice spacing, and $A_{cl}(t)$ is the classical vector potential of the driving laser with polarization in the direction of the chain. The onsite and nearest-neighbor electron-electron interactions are described by the first and second terms in Eq. (\ref{eq:H_ee}), respectively. More details on the Fermi-Hubbard model can be found in, e.g., Ref. \cite{Essler2005}. We restrict ourselves to the case of $U = 12 t_0$ and $V= 4 t_0$, which puts the system in the charge-density phase supporting a so-called Mott exciton. The quantum-optical response of this system is well-studied in the Schrödinger picture in Ref. \cite{Lange2025b} (see also Refs. \cite{Lange2024a, Lange2025a}), which we benchmark against the PHD results presented below. As in earlier work, we use a hopping amplitude of $t_0= 0.0191$ a.u. and a lattice spacing of $a = 7.5589$ a.u. \cite{Silva2018, Hansen2022a, Hansen2022b, Lange2024} and employ periodic boundary conditions of a chain with $L=8$ sites \cite{Lange2024a, Lange2025a, Lange2025b}. We drive the system with a coherent driving pulse $A_{cl}(t) = (F_0 / \omega_L) \sin^2[\omega_L t / (2 N_c)] \sin(\omega_L t + \pi/2)$, with $F_0 = t_0/a = 0.0025$ a.u. corresponding to a peak intensity of $2.2 \times 10^{11}$ W/cm$^2$ and a carrier laser frequency of $\omega_L = t_0/2 = 0.00955$ a.u. $= 396$ THz angular frequency over $N_c = 10$ cycles. The semiclassical dynamics were obtained using the Arnoldi-Lanczos algorithm \cite{Park1986, Smyth1998, Guan2007, Frapiccini2014} with a Krylov subspace of dimension $6$ and a time step size of $dt = 0.26$ a.u., checked for numerical convergence. For the results presented in this strongly-correlated system, a value of $g_0 = 3 \times 10^{-9}$ a.u. is used.

Due to the collective nature of the Hubbard model, we treat each system, i.e., one Hubbard chain with $L=8$ electrons with periodic boundary conditions, as an independent emitter. However, estimating the total number of emitters $N$ is less straightforward than for the atomic ensembles discussed above. We approximate $N$ as follows: first, we estimate how many Hubbard chains can fit on top of each other within the central focus point of a typical laser spot, and then multiply this by the number of sheets that can fit within the thickness of a thin film. In doing so, we make several simplifying assumptions: (i) each sheet experiences the same laser intensity, (ii) propagation effects are neglected, and (iii) all emitters are phase matched.  

Using a beam-spot radius $w_0 = 50~\mu$m, we estimate that the dominant HHG response comes from emitters within $\pm 5~\mu$m along the beam axis, giving an effective height of $10~\mu$m $\approx 2 \times 10^5$ a.u. For a cubic lattice structure, each sheet then contains $N_{\text{sheet}} \approx 2.5 \times 10^4$ Hubbard chains stacked on top of each other with a single Hubbard chain in the horizontal direction. For a film thickness of $t_{\text{film}} = 100$~nm $\approx 2 \times 10^3$ a.u., corresponding to $\approx 2.5 \times 10^2$ sheets, we estimate the total number of independent Hubbard chains as $N \approx 6.3 \times 10^6$. This value can increase substantially for thicker crystals, though the above approximations become less reliable in that regime, as the emitters are unlikely to experience the same driving-field intensity due to absorption.

With this estimate of $N$, we show the relevant observables in Fig.~\ref{fig:fhfullfigure}. For clarity, results are presented for a broad range of numbers of emitters, $N = 10^3, 10^5, 10^7$. The HHG spectra (top row) show that, for all $N$, the coherent contribution dominates, with the spectrum peaking at the energy of the Mott exciton, as discussed in Ref.~\cite{Lange2025b}. The quadrature squeezing (middle row) exhibits the same $N$-scaling as observed for atomic ensembles in Fig.~\ref{fig:atomfullfigure} and predicted by Eq.~(\ref{eq:quadrature_variance_peturbative}). We note a strong signal in the degree of squeezing right at the energy of the Mott exciton, see a discussion in Ref. \cite{Lange2025b}. Interestingly, we have found for two very different electronic systems that a strong resonance gives rise to an increase in the degree of squeezing, indicating a trend. The second-order photon correlation function (bottom row) indicates that deviations from Poissonian statistics only occur for a relatively small number of emitters. In realistic solid-state experiments, where $N$ is significantly larger than in gas-phase setups, the photon statistics are therefore expected to be very close to classical Poissonian statistics. This contrasts with the experimental observations of Ref.~\cite{Theidel2024}. We note, however, that the present study considers a strongly correlated system, whereas Ref.~\cite{Theidel2024} investigates band-gap materials. Further, Ref. \cite{Theidel2024} finds a significant dependence on the laser intensity, indicating that as more emitters are participating in the HHG process, the photon statistics approach classical Poissonian statistics, which is consistent with the findings of this work. A precise determination of $N$ in solid-state targets as a function of the driving-field intensity remains a topic for future investigation.

Looking generally at the nonclassical response from both an atomic ensemble in Fig. \ref{fig:atomfullfigure} and a strongly correlated system in Fig. \ref{fig:fhfullfigure}, we note that the degree of squeezing increases around a resonance in the electronic system with the total squeezing increasing with the number of emitters, $N$. Curiously, we see that the same resonances are not similarly identifiable from the photon statistics, and at the same time, the photon statistics tend towards a classical Poissonian distribution for an increasing number of emitters. This peculiar contrast, that the degree of squeezing shows clear quantum light enhanced by resonances and the number of emitters, while the photon statistics does not show systematically any dependence on resonances and at the same time tending towards a classical distribution, is an interesting topic in itself. While the PHD allows one to investigate the underlying emitter dynamics to resolve this trend, we leave this as a topic for future work.

\subsection{The effect of quantum corrections to semiclassical dynamics} \label{Sec:QuantumCorrectionsToSemiclassical}
\begin{figure}
	\centering
	\includegraphics[width=1\linewidth]{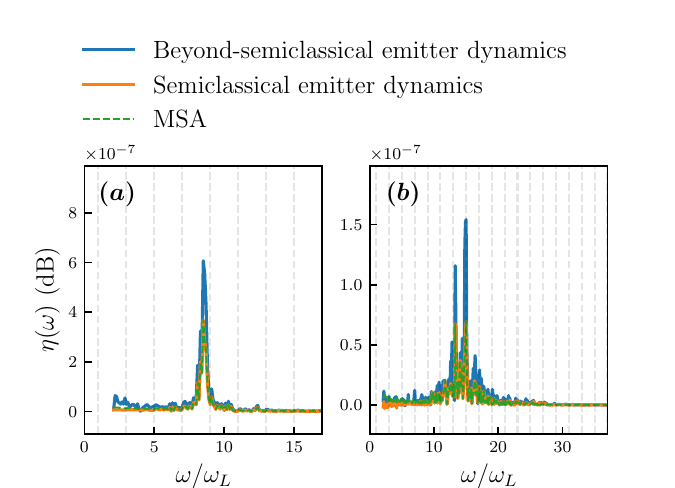}
	\caption{Degree of squeezing from beyond-semiclassical emitter dynamics $\hat{\bm{Q}}'(t) = \hat{\bm{Q}}_{sc}(t) + \hat{\bm{Q}}_q(t)$ [Eqs. (\ref{eq:Q_prime_both_contributions})-(\ref{eq:Q_H_expansion}), solid blue] compared to the effect only semiclassical emitter dynamics $\hat{\bm{Q}}'(t) = \hat{\bm{Q}}_{sc}(t)$ [Eq. (\ref{eq:Q_definition}), solid orange]. These PHD expressions are compared to the MSA prediction (dashed green). (a) Results for an ensemble of atoms. (b) Results for a strongly correlated material using the Fermi-Hubbard model. The results are shown for $N = 10^5$ independent emitters in both cases. The corrections due to the coupling to the quantized electromagnetic field [$\hat{\bm{Q}}_q(t)$, Eqs. (\ref{eq:Q_prime_both_contributions})-(\ref{eq:Q_H_expansion})] significantly contribute to the degree of squeezing and can hence not be neglected. The fundamental mode is not shown as discussed in the main text. See Sec. \ref{Sec:Results} for laser parameters.}
	\label{fig:squeezingdifferences}
\end{figure}

We emphasize that Eq. (\ref{eq:Q_prime_both_contributions}) includes both the semiclassical dynamics for the emitter and the corrections due to the coupling between the emitter and the quantized electromagnetic field [to second order in the coupling, $g_0 N \lvert \tilde{p}(\omega) \rvert$]. Interestingly, both the semiclassical dynamics and beyond-semiclassical dynamics affect the nonclassical nature of the emitted HHG. Here, we investigate the quantitative effect of the beyond-semiclassical dynamics on the degree of squeezing. The quadrature variance relevant for the degree of squeezing is given in Eq. (\ref{eq:quadrature_variance_peturbative}), and, as stated in Sec. \ref{Sec:Perturbative_dynamics}, the last integral of Eq. (\ref{eq:quadrature_variance_peturbative}) is due to the beyond-semiclassical dynamics, while the first two terms of Eq. (\ref{eq:quadrature_variance_peturbative}) originates from semiclassical dynamics. In Fig. \ref{fig:squeezingdifferences}, we calculate the squeezing with and without the beyond-semiclassical emitter dynamics to investigate how much these fluctuations affect the degree of squeezing. Mathematically, this means that we compare the effect of a beyond-semiclassical emitter dynamics $\hat{\bm{Q}}'(t) = \hat{\bm{Q}}_{sc}(t) + \hat{\bm{Q}}_q(t)$ [Eqs. (\ref{eq:Q_prime_both_contributions})-(\ref{eq:Q_H_expansion}), solid blue] with a purely semiclassical description of the emitter dynamics $\hat{\bm{Q}}'(t) = \hat{\bm{Q}}_{sc}(t)$ [Eq. (\ref{eq:Q_definition}), solid orange] on the degree of squeezing. For completeness, we also compare these to the Markov-state Approximation (MSA, dashed green), which is an approximative description of the final quantum state found from a Schrödinger-picture approach, which does not consider beyond-semiclassical emitter dynamics, see Eq. (41) of Ref. \cite{Lange2025a}. The comparison is shown in Fig. \ref{fig:squeezingdifferences}, where we see both for the case of an atomic ensemble [Fig. \ref{fig:squeezingdifferences}(a)] and for the strongly correlated material [Fig. \ref{fig:squeezingdifferences}(b)], that the inclusion of the fluctuations to the quantized electromagnetic field contributes significantly to the degree of squeezing. This is an important finding, as one might be tempted to neglect this correction due to the strong driving field inducing the semiclassical dynamics on the emitter. Further, we note that the MSA prediction closely coincides with the PHD expression using only $\hat{\bm{Q}}_{sc}(t)$, showing that the MSA indeed does not take the quantum-induced fluctuations into account. However, if one is only interested in the HHG spectrum, this quantum correction does not alter the results from semiclassical dynamics, as seen in Eq. (\ref{eq:spectrum_perturbative}), i.e., calculating the harmonic spectrum using semiclassical methods is as accurate as the PHD prediction of the spectrum.

\section{Discussion} \label{Sec:Discussion}

\begin{table*}
	\begin{ruledtabular}
		\begin{tabular}{ p{3cm} p{6cm} p{6cm} }
			{\centering \textbf{Property}\par} 
			& {\centering \textbf{Strong-field quantum optical perturbative Heisenberg dynamics (PHD)}\par}
			& {\centering \textbf{Schrödinger-picture approach}\par} \\
			\hline
			
			{\centering Nature of approximations\par}
			& {\centering Well controlled in orders of the light-matter coupling and collective dipole response. Accurate in regime relevant for HHG experiments\par}
			& {\centering Approximations are not controlled or readily possible to benchmark\par} \\
			
			{\centering Description of emitter dynamics\par}
			& {\centering Semiclassical dynamics with corrections due to coupling to the quantized field [Eqs. (\ref{eq:Q_prime_both_contributions})-(\ref{eq:Q_H_expansion})]\par}
			& {\centering Only semiclassical dynamics\par} \\
			
			{\centering Origin of photonic dynamics\par}
			& {\centering Due to both semiclassical and \textit{quantum-induced} emitter correlations [Eq. (\ref{eq:Heisenberg_perturbative})] \par}
			& {\centering Only due to semiclassical emitter correlations\par} \\
			
			{\centering  Scaling of observables with number of emitters ($N$)\par}
			& {\centering Follows naturally for all considered observables\par}
			& {\centering Is not included naturally for any observable\par} \\
			
			{\centering Link between photonic observables and emitter dynamics \par}
			& {\centering Appears explicitly in the PHD expression for the observables [Eqs. (\ref{eq:spectrum_perturbative}), (\ref{eq:quadrature_variance_peturbative}), and (\ref{eq:g2_perturbative})]\par}
			& {\centering Requires further approximations (for example MSA) \par} \\
			
			{\centering Numerical cost in addition to solving the semiclassical TDSE \par}
			& {\centering Simple integration of the time-dependent dipole and its time correlations [See Eqs. (\ref{eq:spectrum_perturbative}), (\ref{eq:quadrature_variance_peturbative}), and (\ref{eq:g2_perturbative})] \par}
			& {\centering Integration of a large set of coupled differential equations for the photonic states [Eq. (\ref{eq:chi_EOM_schrodinger})]\par} \\
		\end{tabular}
	\end{ruledtabular}
	\caption{Comparison of the strong-field quantum optical PHD to the Schrödinger-picture approach for quantum optical HHG. The properties listed in the table regarding the Schrödinger-picture approach can be found in, e.g., Ref \cite{Lange2025a}.}
	\label{tab:comparison}
\end{table*}
\subsection{Advantages of the PHD compared to the Schrödinger picture} \label{Sec:Dis_AdvantagesPHD}
As discussed in Sec.~\ref{Sec:Introduction}, the PHD offers several advantages compared to the Schrödinger-picture approach. While we do not explain it in great detail, we make general comments on the Schrödinger-picture approach encapsulated by the literature, see, e.g., Refs. \cite{Gorlach2020, Lewenstein2021, Stammer2023, Stammer2024a, Lange2024a, Yi2024, Rivera-Dean2024d, Lange2025a}. First, we emphasize that, as it is not possible to obtain a closed-form expression for either electron or photon dynamics using either the Heisenberg or Schrödinger picture, approximations are required. While the PHD builds on a controlled and extendable perturbative expansion in the effective coupling, $g_0 N \lvert \tilde{p}(\omega) \rvert$, the typically approximations in the Schrödinger picture, such as the decoupling of photonic modes, where all photonic modes are treated independently in a product ansatz for the photonic wavefunction, is neither controlled nor similarly possible to iterate, in order to obtain a more accurate description \cite{Gorlach2020, Lange2024a, Lange2025a}. As such, the PHD expressions for observables are both easier to extend for higher accuracy and rely on a better physical understanding. Further, the PHD provides a beyond-semiclassical description of the electron dynamics in strong-field physics as seen from Eq. (\ref{eq:Q_prime_both_contributions}) and illustrated in Fig. \ref{fig:dipolephotoninteraction}. In general, if one wants to keep all photonic modes and transition dipole matrix elements, the inclusion of quantum-induced corrections to semiclassical emitter dynamics is not possible due to the approximations employed in the Schrödinger-picture approach. Thus, to the best of our knowledge, the beyond-semiclassical emitter dynamics captured by the PHD have not been accounted for previously in strong-field quantum optics. Consequently, the photonic observables of interest, i.e., the harmonic spectrum, the degree of squeezing, and photon statistics, are modified by these quantum-induced corrections to semiclassical emitter dynamics as discussed in Sec. \ref{Sec:QuantumCorrectionsToSemiclassical} above. Interestingly, the leading-order term in the PHD expansion does not couple different frequency modes, thereby justifying the mode-decoupling ansatz used in the Schrödinger picture \cite{Gorlach2020, Lange2024a, Lange2025a, Lange2025b}—an assumption that cannot be benchmarked internally within the Schrödinger approach. However, the Schrödinger-picture results for the Fermi–Hubbard model in Refs. \cite{Lange2024a, Lange2025a, Lange2025b} capture only the coherent contribution to the spectrum, despite not explicitly considering the number of emitters. This limitation is an unintended consequence of the mode-decoupling ansatz and is discussed in detail in App. \ref{App:product_ansatz_consequence}.

An additional advantage of the PHD is the fact that it naturally shows how each observable scales with the number of independent emitters, $N$. Interestingly, we find that while the degree of squeezing increases with $N$, the photon statistics tend towards a classical Poissonian distribution as $N$ grows, see Figs. \ref{fig:atomfullfigure} and \ref{fig:fhfullfigure}. Such findings are not obtainable in the Schrödinger picture as $N$ is not naturally introduced in the formalism. Another advantage of the PHD is the fact that the expressions for the observables of interest [Eqs. (\ref{eq:spectrum_perturbative}), (\ref{eq:quadrature_variance_peturbative}), and (\ref{eq:g2_perturbative})], contain a clear link to the underlying electron dynamics, in particular revealing the origin of the nonclassical response of the system. Such an insight is only obtainable in the Schrödinger picture with additional approximations, such as the Markov-state approximation (MSA), see Refs. \cite{Stammer2024a, Lange2025a}. Finally, the numerical cost required to obtain the expectation values of interest is much lower in the PHD as only a simple integration has to be performed, compared to the Schrödinger picture, where a generally large set of coupled differential equations has to be integrated, which is numerically more taxing. We have summarized the differences between the PHD and the Schrödinger-picture approaches in strong-field quantum optics in Tab. \ref{tab:comparison} for clarity.

\subsection{How squeezed is the light actually?}
As shown in Figs. \ref{fig:atomfullfigure} and \ref{fig:fhfullfigure}, a realistic number of independent, phase-matched emitters (under the assumptions in Sec. \ref{Sec:Results}) yields a degree of squeezing $\eta_{\bm{k},\sigma} \ll 1$dB. While many quantum-optical platforms routinely measure squeezing levels of $\eta \simeq 1$–$10$ dB at wavelengths $\lambda > 800$ nm \cite{Andersen2016}, the results presented here—together with other works in strong-field quantum optics—demonstrate a route to generating squeezed light at significantly shorter wavelengths, extending into the UV and XUV regimes. Thus, although the achievable squeezing is currently modest compared to standard experiments \cite{Andersen2016}, strong-field quantum optics provides a means of probing nonclassical light in previously inaccessible spectral regions.

Enhancing the squeezing would require emitter systems with stronger nonclassical responses, for example, via strong resonances in the electronic system, as indicated in Figs. \ref{fig:atomfullfigure} and \ref{fig:fhfullfigure}, or experimental designs that increase the light–matter coupling, for example via cavity enhancement \cite{Yi2024, Klimkin2025}. Such improvements may enable applications in quantum information science and quantum metrology \cite{Lewenstein2024AttoAndQI, Degen2017}. Nevertheless, we emphasize that the nonclassical response remains a valuable spectroscopic probe of the electronic medium, regardless of the absolute squeezing level, provided it lies above the experimental noise floor.

\subsection{The role of the transition dipoles} \label{Sec:Disc_Role_Of_Transition_Dipoles}
The nonclassical observables depend on the correlations of the induced emitter dynamics. These correlations can be reformulated via the transition dipoles (or sometimes referred to as transition currents) as seen in Eq. (\ref{eq:correlations_to_cross_dipoles}) and it follows from Eqs. (\ref{eq:quadrature_variance_peturbative}) and (\ref{eq:g2_perturbative}) that the off-diagonal elements $p_{m,n}^{(\sigma)}(t)$ ($m\neq i$ and $n \neq i$) are required to generate the nonclassical response in HHG; retaining only diagonal terms is equivalent to neglecting the time correlations which then produces purely coherent radiation (see also Apps. \ref{App:Photon_statistics} and \ref{App:Exp_electron_operators}). 

In the Schrödinger picture, the equation of motion for the photonic state $\ket{\chi^{(m)}(t)}$ associated with the emitter eigenstate $\ket{\phi_m}$ is \cite{Lange2025a}
\begin{equation}
	i \partial_t \ket{\chi^{(m)}(t)} 
	= \hat{\bm{A}}(t)\cdot \sum_n \bm{p}_{m,n}(t)\ket{\chi^{(n)}(t)}, 
	\label{eq:chi_EOM_schrodinger}
\end{equation}
where the coupling to other photonic states is weighted by these transition dipoles. Although some models (e.g., one-band systems) allow the off-diagonal dipoles to be neglected exactly, this is generally not the case \cite{Lange2024a}. Nonetheless, many works omit them by appealing to negligible ground-state depletion~\cite{Lewenstein2021, Rivera-Dean2022, Stammer2023, Bhattacharya2023, Stammer2024, Stammer2025b}. The PHD, on the other hand, highlights that transition dipoles represent fluctuations and time correlations of the emitter, not population transfer to excited states, as explicitly shown in Eq. (\ref{eq:correlations_to_cross_dipoles}). Hence, the off-diagonal elements enter solely as a compact representation of emitter correlations; they do not imply population of excited states. In fact, many of these off-diagonal transition dipole elements can be orders of magnitude larger than the diagonal contributions \cite{Gorlach2020, Lange2024a}.

A simple example illustrates this point. For a general inversion-symmetric system without external driving, one has $\bm{p}_{i,i}(t)=0$, while $\bm{p}_{i,m}(t)= e^{-i(E_i-E_m)t}\bm{p}_{i,m}(0)\neq 0$.
The off-diagonal elements therefore, remain finite even when the emitter stays entirely in its ground state; they merely encode its intrinsic time correlations, which, in the absence of a driving field, vanish only after time integration. Thus, one cannot discard the transition-dipole elements on the basis of small excited-state population, or, equivalently, negligible ground-state depletion, as they fundamentally quantify fluctuations of the initial state. 

\section{Conclusion and outlook} \label{Sec:Conlcusion}

In this work, we derived closed-form expressions for photonic operators associated with matter driven by intense laser fields in the strong-field quantum optical PHD formalism. These expressions were obtained via a perturbative expansion in the weak light-matter coupling parameter after transferring the driving field from the initial state to the Hamiltonian via unitary transformations, yielding controlled and systematically extendable approximations. In turn, these expressions include corrections to the semiclassical emitter dynamics due to the coupling to the quantized electromagnetic field, which consequently notably affects the photonic observables. Assuming identical and uncorrelated emitters, we characterized how relevant observables scale with the number of independent emitters, $N$: the spectrum shows the expected $N^2$ (coherent) and $N$ (incoherent) scaling; the quadrature variance relevant to squeezing scales linearly with $N$; and the second-order correlation function $g^{(2)}_{\bm{k}, \sigma}(0)$ approaches Poissonian statistics for sufficiently large $N$. The strong-field quantum optical PHD framework further makes the microscopic origin of the nonclassical response transparent and has a lower computational cost than similar approaches relying on the Schrödinger picture for calculating photonic observables.

We evaluated the PHD expressions for both an atomic ensemble and a strongly correlated material modeled via the Fermi–Hubbard Hamiltonian, illustrating the generality of the approach. In both cases, the results for the harmonic spectra agree with those obtained in the Schrödinger picture under standard approximations, validating their use in this case. However, accurate prediction of the degree of squeezing of the emitted light requires the beyond-semiclassical emitter dynamics captured by PHD, which is missing in the Schrödinger-picture approach. Given these advantages—and the reduced numerical complexity—we find the PHD approach preferable to Schrödinger-picture treatments in many contexts.

For fully quantitative predictions, however, additional effects must be incorporated, including propagation of the generated light through the medium and decoherence in the electronic system. While the PHD offers an efficient and controlled method for computing photonic observables, a more complete treatment of the light-propagation and matter-dynamics aspects remains an important task for future work.

	\begin{acknowledgements}
		This work is supported by the Independent Research Fund Denmark (Technology and Production Sciences 10.46540/4286-00053B) and the Novo Nordisk Foundation Project Grants in the Natural and Technical Sciences (0094623).
	\end{acknowledgements}
	
	\appendix
	\onecolumngrid
	\allowdisplaybreaks 
\section{Photon statistics in exact dynamics and PHD} \label{App:Photon_statistics}

In this appendix, we derive and provide the full expressions for the $g_{\bm{k}, \sigma}^{(2)}(0)$ in both the exact case [Eq. (\ref{eq:g2_exact})] and in the PHD [Eq. (\ref{eq:g2_perturbative})]. For convenience, we write out the exact expression for the second order correlation function, restating Eq. (\ref{eq:g2_general})
\begin{equation}
	g_{\bm{k}, \sigma}^{(2)}(0) = \frac{\langle \hat{a}_{\bm{k}, \sigma}^\dagger \hat{a}_{\bm{k}, \sigma}^\dagger \hat{a}_{\bm{k}, \sigma} \hat{a}_{\bm{k}, \sigma} \rangle}{ \langle \hat{a}_{\bm{k}, \sigma}^\dagger \hat{a}_{\bm{k}, \sigma} \rangle^2},
\end{equation}
whose general form must be, conferring with Appendix \ref{App:Exp_electron_operators},
\begin{equation}
	g_{\bm{k}, \sigma}^{(2)}(0) = \dfrac{\dfrac{N!}{(N-4)!} C^{(0)}_{\bm{k}, \sigma} + \dfrac{N!}{(N-3)!} C^{(2)}_{\bm{k}, \sigma} + \dfrac{N!}{(N-2)!} (C^{(2,2)}_{\bm{k}, \sigma} + C^{(3)}_{\bm{k}, \sigma}) + \dfrac{N!}{(N-1)!} C^{(4)}_{\bm{k}, \sigma}}{N^2(N-1)^2 C^{(0)}_{\bm{k}, \sigma} + 2N^2 (N-1) \sqrt{C^{(0)}_{\bm{k}, \sigma}} \sqrt{\tilde{C}^{(2)}_{\bm{k}, \sigma}} + N^2 \tilde{C}^{(2)}_{\bm{k}, \sigma}}, \label{eq:g2_exactform}
\end{equation}
where the terms have been ordered according to their $N$-dependence.
We first note that the expression in the denominator is simply the counting operator (squared), which has already been expressed for the spectrum, yielding, 
\begin{align}
	\langle \hat{a}_{\bm{k}, \sigma}^\dagger (t) \hat{a}_{\bm{k}, \sigma} (t) \rangle &= \dfrac{g_0^2}{\omega_k} \int_{0}^{t} dt_1  \int_{0}^{t} dt_2 \langle \hat{\bm{e}}_\sigma \cdot \hat{\bm{Q}}(t_1) \hat{\bm{e}}_\sigma \cdot \hat{\bm{Q}}(t_2) \rangle e^{-i\omega_k(t_1-t_2)} \nonumber \\
	&= \dfrac{g_0^2}{\omega_k} \left( N \int_{0}^{t} dt_1  \int_{0}^{t} dt_2 \langle \hat{{p}}_\sigma(t_1)  \hat{{p}}_\sigma(t_2) \rangle e^{i\omega_k(t_1-t_2)} + N(N-1) \bigg \lvert \int_0^t dt_1 e^{-i \omega_k t_1} \langle  \hat{{p}}_\sigma(t_1) \rangle \bigg \rvert^2 \right), 
	\label{eq:counting_operator_exact}
\end{align}
where $ \hat{{p}}_\sigma(t) = \hat{\bm{p}} \cdot \hat{\bm{e}}_\sigma$. Here, the first term of Eq. (\ref{eq:Heisenberg_solution_exact}) of the main text does not contribute to the final expression, since $\hat{a}_{\bm{k}, \sigma} (0)$ [$\hat{a}_{\bm{k}, \sigma}^\dagger (0)$] will annihilate the initial state when the operator acts on it from the left (right) as all modes are in the vacuum state initially. Inserting the exact expression for $\hat{a}_{\bm{k}, \sigma}(t)$ from Eq. (\ref{eq:Heisenberg_solution_exact}) in the numerator of Eq. (\ref{eq:g2_exactform}) yields an expression that does not simplify as easily, since $[\hat{\bm{e}}_\sigma \cdot \hat{\bm{Q}}(t), \hat{a}_{\bm{k}, \sigma} (0)] \neq 0$. We obtain
\begin{align}
	\langle \hat{a}_{\bm{k}, \sigma}^\dagger (t) \hat{a}_{\bm{k}, \sigma}^\dagger (t) \hat{a}_{\bm{k}, \sigma} (t) \hat{a}_{\bm{k}, \sigma} (t)\rangle &= \dfrac{g_0^4}{\omega_k^2} \int_{0}^{t} dt_1  \int_{0}^{t} dt_2 \int_{0}^{t} dt_3  \int_{0}^{t} dt_4  e^{-i\omega_k(t_1 + t_2 - t_3 - t_4)} \nonumber \\
	& \qquad\qquad\qquad \langle \hat{\bm{e}}_\sigma \cdot \hat{\bm{Q}}(t_1) \hat{\bm{e}}_\sigma \cdot \hat{\bm{Q}}(t_2) \hat{\bm{e}}_\sigma \cdot \hat{\bm{Q}}(t_3) \hat{\bm{e}}_\sigma \cdot \hat{\bm{Q}}(t_4) \rangle \nonumber
	\\&+ \dfrac{g_0^2}{\omega_k} \int_{0}^{t} dt_1  \int_{0}^{t} dt_2 \langle \hat{\bm{e}}_\sigma \cdot \hat{\bm{Q}}(t_1) \hat{a}_{\bm{k}, \sigma}^\dagger (0)\hat{a}_{\bm{k}, \sigma}(0)\hat{\bm{e}}_\sigma \cdot \hat{\bm{Q}}(t_2) \rangle e^{-i\omega_k(t_1 - t_2)},
\intertext{
from which the $N$-dependence can be explicitly written out in accordance with App. \ref{App:Exp_electron_operators}. This yields}
	\langle \hat{a}_{\bm{k}, \sigma}^\dagger (t) \hat{a}_{\bm{k}, \sigma}^\dagger (t) \hat{a}_{\bm{k}, \sigma} (t) \hat{a}_{\bm{k}, \sigma} (t)\rangle 
	&= \dfrac{g_0^4}{\omega_k^2} \int_{0}^{t} dt_1  \int_{0}^{t} dt_2 \int_{0}^{t} dt_3  \int_{0}^{t} dt_4 e^{-i\omega_k(t_1 + t_2 - t_3 - t_4)} \nonumber \\
	& \qquad \bigg[ N\langle \hat{{p}}_\sigma(t_1) \hat{{p}}_\sigma(t_2) \hat{{p}}_\sigma(t_3) \hat{{p}}_\sigma(t_4) \rangle  \nonumber \\
	&\qquad+\dfrac{N!}{(N-2)!}(2\langle \hat{{p}}_\sigma(t_1) \hat{{p}}_\sigma(t_2) \hat{{p}}_\sigma(t_3) \rangle\langle \hat{{p}}_\sigma(t_4) \rangle 
	+2\langle \hat{{p}}_\sigma(t_1) \hat{{p}}_\sigma(t_3) \hat{{p}}_\sigma(t_4) \rangle\langle \hat{{p}}_\sigma(t_2) \rangle )  \nonumber \\
	&\qquad+\dfrac{N!}{(N-2)!} \left( \langle \hat{{p}}_\sigma(t_1) \hat{{p}}_\sigma(t_2) \rangle\langle \hat{{p}}_\sigma(t_3) \hat{{p}}_\sigma(t_4) \rangle + 2\langle \hat{{p}}_\sigma(t_1) \hat{{p}}_\sigma(t_3) \rangle \langle \hat{{p}}_\sigma(t_2)  \hat{{p}}_\sigma(t_4) \rangle \right) \nonumber \\
	&\qquad+ \dfrac{N!}{(N-3)!}( \langle \hat{{p}}_\sigma(t_1) \hat{{p}}_\sigma(t_2) \rangle\langle \hat{{p}}_\sigma(t_3) \rangle\langle \hat{{p}}_\sigma(t_4) \rangle 
	+ \langle \hat{{p}}_\sigma(t_3) \hat{{p}}_\sigma(t_4) \rangle\langle \hat{{p}}_\sigma(t_1) \rangle\langle \hat{{p}}_\sigma(t_2) \rangle 
	\nonumber \\
	&\qquad\qquad\qquad\qquad\qquad+ 4\langle \hat{{p}}_\sigma(t_1) \hat{{p}}_\sigma(t_3) \rangle \langle \hat{{p}}_\sigma(t_2) \rangle \langle \hat{{p}}_\sigma(t_4) \rangle ) \nonumber \\
	&\qquad+ \dfrac{N!}{(N-4)!} \langle \hat{{p}}_\sigma(t_1) \rangle\langle \hat{{p}}_\sigma(t_2) \rangle\langle \hat{{p}}_\sigma(t_3) \rangle\langle \hat{{p}}_\sigma(t_4) \rangle \bigg] \nonumber
	\\& + \dfrac{g_0^2}{\omega_k} \bigg[ N \int_{0}^{t} dt_1  \int_{0}^{t} dt_2 e^{-i\omega_k(t_1 - t_2)} \langle \hat{{p}}_\sigma(t_1) \hat{a}_{\bm{k}, \sigma}^\dagger (0) \hat{a}_{\bm{k}, \sigma}(0)\hat{{p}}_\sigma(t_2) \rangle   \nonumber \\
	&\qquad\qquad\qquad\qquad\qquad+\dfrac{N!}{(N-2)!} \bigg \lvert \int_0^t dt_1 e^{-i \omega_k t_1} \langle \hat{a}_{\bm{k}, \sigma}(0)\hat{{p}}_\sigma(t_1)\rangle \bigg \rvert^2 \bigg]. 
\end{align}
The expression above has been simplified utilizing the pairwise equivalence between $t_1$ and $t_2$ and $t_3$ and $t_4$. Recasting the expression into the form of Eq. (\ref{eq:g2_exactform}) gives an exact expression for the coefficients,
\begin{subequations}
\begin{align}
	C^{(0)}_{\bm{k}, \sigma} &= \dfrac{g_0^4}{\omega_k^2} \int_{0}^{t} dt_1  \int_{0}^{t} dt_2 \int_{0}^{t} dt_3  \int_{0}^{t} dt_4 e^{-i\omega_k(t_1 + t_2 - t_3 - t_4)}\langle \hat{{p}}_\sigma(t_1) \rangle\langle \hat{{p}}_\sigma(t_2) \rangle\langle \hat{{p}}_\sigma(t_3) \rangle\langle \hat{{p}}_\sigma(t_4) \rangle \nonumber \\
	&= \dfrac{g_0^4}{\omega_k^2} \bigg \lvert \int_0^t dt_1 e^{-i \omega_k t_1} \langle  \hat{{p}}_\sigma(t_1) \rangle \bigg \rvert^4 \\
	C^{(2)}_{\bm{k}, \sigma} &= \dfrac{g_0^4}{\omega_k^2} \int_{0}^{t} dt_1  \int_{0}^{t} dt_2 \int_{0}^{t} dt_3  \int_{0}^{t} dt_4 e^{-i\omega_k(t_1 + t_2 - t_3 - t_4)} \bigg[ \langle \hat{{p}}_\sigma(t_1) \hat{{p}}_\sigma(t_2) \rangle\langle \hat{{p}}_\sigma(t_3) \rangle\langle \hat{{p}}_\sigma(t_4) \rangle \nonumber \\
	&\qquad\qquad\qquad\qquad\qquad
	+ \langle \hat{{p}}_\sigma(t_3) \hat{{p}}_\sigma(t_4) \rangle\langle \hat{{p}}_\sigma(t_1) \rangle\langle  \hat{{p}}_\sigma(t_2) \rangle + 4\langle \hat{{p}}_\sigma(t_1) \hat{{p}}_\sigma(t_3) \rangle \langle \hat{{p}}_\sigma(t_2) \rangle \langle \hat{{p}}_\sigma(t_4) \rangle \bigg] \nonumber \\
	&= \dfrac{g_0^4}{\omega_k^2} \bigg\{ 2\text{Re}\left[\int_{0}^{t} dt_1  \int_{0}^{t} dt_2 e^{-i\omega_k(t_1+t_2)} \langle \hat{{p}}_\sigma(t_1)  \hat{{p}}_\sigma(t_2) \rangle \cdot  \left( \int_0^t dt_3 e^{i \omega_k t_3} \langle  \hat{{p}}_\sigma(t_3) \rangle \right)^2\right]  \nonumber \\
	&\qquad\qquad\qquad\qquad\qquad
	+ 4\int_{0}^{t} dt_1  \int_{0}^{t} dt_3 e^{-i\omega_k(t_1-t_3)} \langle \hat{{p}}_\sigma(t_1)  \hat{{p}}_\sigma(t_3) \rangle \cdot  \bigg \lvert \int_0^t dt_2 e^{-i \omega_k t_2} \langle  \hat{{p}}_\sigma(t_2) \rangle \bigg \rvert^2 \bigg\}  \\
	C^{(3a)}_{\bm{k}, \sigma} &= \dfrac{g_0^4}{\omega_k^2} \int_{0}^{t} dt_1  \int_{0}^{t} dt_2 \int_{0}^{t} dt_3  \int_{0}^{t} dt_4 e^{-i\omega_k(t_1 + t_2 - t_3 - t_4)}\nonumber \\
	&\qquad\qquad\qquad\qquad\qquad
	\bigg[2\langle \hat{{p}}_\sigma(t_1) \hat{{p}}_\sigma(t_2) \hat{{p}}_\sigma(t_3) \rangle\langle \hat{{p}}_\sigma(t_4) \rangle 
	+2\langle \hat{{p}}_\sigma(t_1) \hat{{p}}_\sigma(t_3) \hat{{p}}_\sigma(t_4) \rangle\langle \hat{{p}}_\sigma(t_2) \rangle  \nonumber \\
	&\qquad\qquad\qquad\qquad\qquad
	+ \langle \hat{{p}}_\sigma(t_1) \hat{{p}}_\sigma(t_2) \rangle\langle \hat{{p}}_\sigma(t_3) \hat{{p}}_\sigma(t_4) \rangle + 2\langle \hat{{p}}_\sigma(t_1) \hat{{p}}_\sigma(t_3) \rangle \langle \hat{{p}}_\sigma(t_2)  \hat{{p}}_\sigma(t_4) \rangle\bigg]  \nonumber\\
	&= \dfrac{g_0^4}{\omega_k^2} \bigg\{ 4\text{Re}\left[ \int_{0}^{t} dt_1  \int_{0}^{t} dt_2 \int_{0}^{t} dt_3  \int_{0}^{t} dt_4 e^{-i\omega_k(t_1 + t_2 - t_3 - t_4)} \langle \hat{{p}}_\sigma(t_1) \hat{{p}}_\sigma(t_3) \hat{{p}}_\sigma(t_4) \rangle\langle \hat{{p}}_\sigma(t_2) \rangle  \right] \nonumber \\
	&\qquad\qquad\qquad\qquad\qquad
	+\bigg \lvert \int_{0}^{t} dt_1  \int_{0}^{t} dt_2 e^{-i\omega_k(t_1+t_2)} \langle \hat{{p}}_\sigma(t_1)  \hat{{p}}_\sigma(t_2) \rangle \bigg \rvert^2 \nonumber \\
	&\qquad\qquad\qquad\qquad\qquad
	+\bigg[ \int_{0}^{t} dt_1  \int_{0}^{t} dt_2 e^{i\omega_k(t_1-t_2)} \langle \hat{{p}}_\sigma(t_1)  \hat{{p}}_\sigma(t_2) \rangle \bigg]^2 \bigg\}  \\
	C^{(3b)}_{\bm{k}, \sigma} &= \dfrac{g_0^2}{\omega_k} \bigg \lvert \int_0^t dt_1 e^{-i \omega_k t_1} \langle \hat{a}_{\bm{k}, \sigma}(0)\hat{{p}}_\sigma(t_1) \rangle \bigg \rvert^2   \\
	C^{(4a)}_{\bm{k}, \sigma} &= \dfrac{g_0^4}{\omega_k^2} \int_{0}^{t} dt_1  \int_{0}^{t} dt_2 \int_{0}^{t} dt_3  \int_{0}^{t} dt_4 e^{-i\omega_k(t_1 + t_2 - t_3 - t_4)} \langle \hat{{p}}_\sigma(t_1) \hat{{p}}_\sigma(t_2) \hat{{p}}_\sigma(t_3) \hat{{p}}_\sigma(t_4) \rangle   \\
	C^{(4b)}_{\bm{k}, \sigma} &= \dfrac{g_0^2}{\omega_k} \int_{0}^{t} dt_1  \int_{0}^{t} dt_2 e^{-i\omega_k(t_1 - t_2)} \langle \hat{{p}}_\sigma(t_1) \hat{a}_{\bm{k}, \sigma}^\dagger (0) \hat{a}_{\bm{k}, \sigma}(0)\hat{{p}}_\sigma(t_2) \rangle  \\
	\tilde{C}^{(2)}_{\bm{k}, \sigma} &= \dfrac{g_0^4}{\omega_k^2} \bigg[\int_{0}^{t} dt_1  \int_{0}^{t} dt_2 e^{i\omega_k(t_1-t_2)}\langle \hat{{p}}_\sigma(t_1)  \hat{{p}}_\sigma(t_2) \rangle \bigg]^2.
\end{align}
\label{eq:g2_exact_expansioncoefficients}
\end{subequations}
Using Eq. (\ref{eq:g2_exact_expansioncoefficients}) in Eq. (\ref{eq:g2_exactform}) yields the full expression in Eq. (\ref{eq:g2_exact}) and concludes the treatment of $g_{\bm{k}, \sigma}^{(2)}(0)$ in the exact formulation.

We now proceed to the derivation of the PHD expression for $g_{\bm{k}, \sigma}^{(2)}(0)$ given in Eq. (\ref{eq:g2_perturbative}). We insert Eq. (\ref{eq:Heisenberg_perturbative}) into Eq. (\ref{eq:g2_general}) and expand the numerator and denominator of $g_{\bm{k}, \sigma}^{(2)}(0)$ up to order $g_0^4N^4$. 

First consider Eq. (\ref{eq:Heisenberg_perturbative}) on the form 
\begin{align}
	\hat{a}_{\bm{k}, \sigma}'(t)  &\simeq \hat{T}^{(0)}_{\bm{k}, \sigma}(t) + g_0 \hat{T}^{(1)}_{\bm{k}, \sigma}(t)
	+ g_0^2 \hat{T}^{(2)}_{\bm{k}, \sigma}(t)
	+ g_0^3 (\hat{T}^{(3a)}_{\bm{k}, \sigma}(t)	
	+ \hat{T}^{(3b)}_{\bm{k}, \sigma}(t)) \\
	\label{eq:terms_of_a_perturbative}
	\hat{T}^{(0)}_{\bm{k}, \sigma}(t) &= \hat{a}'_{\bm{k}, \sigma}(0) e^{-i \omega_k t}\nonumber\\
	\hat{T}^{(1)}_{\bm{k}, \sigma}(t) &= - i \dfrac{1}{\sqrt{\omega_k}} \int_{0}^{t} dt' \hat{Q}_{sc, \sigma}(t') e^{-i\omega_k(t-t')}\nonumber\\
	\hat{T}^{(2)}_{\bm{k}, \sigma}(t) &= \dfrac{1}{g_0\sqrt{\omega_k}} \int_0^t dt' e^{-i \omega_k (t-t')} \int_0^{t'} dt'' [\hat{V}(t''), \hat{Q}_{sc, \sigma}(t')] \nonumber \\
	\hat{T}^{(3a)}_{\bm{k}, \sigma}(t) &=-i \dfrac{1}{g_0^2\sqrt{\omega_k}} \int_0^t dt' e^{-i\omega_k (t-t')} \int_0^{t'} dt'' \int_0^{t'}dt''' \hat{V}(t'') \hat{Q}_{sc, \sigma}(t') \hat{V}(t''') \nonumber \\
	\hat{T}^{(3b)}_{\bm{k}, \sigma}(t) &= i \dfrac{1}{g_0^2 \sqrt{\omega_k}} \int_0^t dt' e^{-i\omega_k(t-t')} \int_0^{t'} dt'' \int_{0}^{t''} dt'''  \hat{Q}_{sc, \sigma}(t') \hat{V}(t'') \hat{V}(t''') + \hat{V}(t''') \hat{V}(t'')  \hat{Q}_{sc, \sigma}(t'),
\end{align}
where the dependence of each term on $g_0$ has been written out explicitly, recalling that each $\hat{V}(t)$ brings a factor of $g_0$. Note that $\hat{T}^{(0)}_{\bm{k}, \sigma}(t)$ holds only photonic operators, while $\hat{T}^{(1)}_{\bm{k}, \sigma}(t)$ holds only semiclassical electronic operators, wherefore the two commute. Furthermore, $\hat{T}^{(2)}_{\bm{k}, \sigma}(t)$ holds one photonic operator, while $\hat{T}^{(3a)}_{\bm{k}, \sigma}(t)$ and $\hat{T}^{(3b)}_{\bm{k}, \sigma}(t)$ each holds two. 
Seeing from this that both $\hat{T}^{(0)}_{\bm{k}, \sigma}(t)$ and $\hat{T}^{(0)}_{\bm{k}, \sigma}(t)\hat{T}^{(1)}_{\bm{k}, \sigma}(t)$ annihilate the initial state when acting on it to the right, and insisting that we should have an even number of photonic operators to obtain a nonvanishing contribution to our expectation value, we find the contributions to the denominator of $g_{\bm{k}, \sigma}^{(2)}(0)$ to be 
\begin{align}
	\langle \hat{a}_{\bm{k}, \sigma}'^\dagger (t) \hat{a}_{\bm{k}, \sigma}'^\dagger (t) \hat{a}_{\bm{k}, \sigma}' (t) \hat{a}_{\bm{k}, \sigma}' (t)\rangle 
	&\simeq g_0^4 \bigg( \langle \hat{T}^{(1)\dagger}_{\bm{k}, \sigma}(t) \hat{T}^{(1)\dagger}_{\bm{k}, \sigma}(t) \hat{T}^{(1)}_{\bm{k}, \sigma}(t)\hat{T}^{(1)}_{\bm{k}, \sigma}(t) \rangle 
	+ \langle \hat{T}^{(2)\dagger}_{\bm{k}, \sigma}(t) \hat{T}^{(0)\dagger}_{\bm{k}, \sigma}(t) \hat{T}^{(0)}_{\bm{k}, \sigma}(t) \hat{T}^{(2)}_{\bm{k}, \sigma}(t) \rangle\nonumber \\
	&\qquad\qquad\qquad\qquad
	+ 2\text{Re}\left\{ \langle \hat{T}^{(1)\dagger}_{\bm{k}, \sigma}(t) \hat{T}^{(1)\dagger}_{\bm{k}, \sigma}(t) \hat{T}^{(0)}_{\bm{k}, \sigma}(t) \hat{T}^{(2)}_{\bm{k}, \sigma}(t) \rangle \right\}
	\bigg),
\end{align}
where higher orders of $g_0$ have been neglected. 

Inserting $\hat{V}(t) = \hat{\bm{A}}(t) \cdot \hat{\bm{Q}}_{sc}(t)$ along with Eq. (\ref{eq:A_operator_2}) allows us to factorize electronic and photonic expectation values, and resolve the latter using $ \langle \hat{a}'_{\bm{k}, \sigma}(0) \hat{a}_{\bm{q}, \lambda}^\dagger (0)\rangle = \delta_{\bm{k}, \bm{q}} \delta_{\sigma, \lambda}$, while all other combinations vanish. We obtain
\begin{align}
	\langle \hat{T}^{(1)\dagger}_{\bm{k}, \sigma}(t) \hat{T}^{(1)\dagger}_{\bm{k}, \sigma}(t) \hat{T}^{(1)}_{\bm{k}, \sigma}(t)\hat{T}^{(1)}_{\bm{k}, \sigma}(t) \rangle 
	&= \dfrac{1}{\omega_k^2}  \int_{0}^{t} dt_1  \int_{0}^{t} dt_2 \int_{0}^{t} dt_3  \int_{0}^{t} dt_4 e^{-i\omega_k(t_1 + t_2 - t_3 - t_4)} \nonumber \\
	&\qquad\qquad\qquad\qquad
	\langle \hat{Q}_{sc, \sigma}(t_1)\hat{Q}_{sc, \sigma}(t_2)\hat{Q}_{sc, \sigma}(t_3)\hat{Q}_{sc, \sigma}(t_4) \rangle   \\
	\langle \hat{T}^{(1)\dagger}_{\bm{k}, \sigma}(t) \hat{T}^{(1)\dagger}_{\bm{k}, \sigma}(t) \hat{T}^{(0)}_{\bm{k}, \sigma}(t) \hat{T}^{(2)}_{\bm{k}, \sigma}(t) \rangle
	&= \dfrac{1}{g_0\omega_k\sqrt{\omega_k}} \int_{0}^{t} dt_1  \int_{0}^{t} dt_2 \int_0^t dt_3 e^{-i\omega_k(t_1 + t_2 - t_3)}
	\int_0^{t_3} dt_4\nonumber \\
	&\qquad\qquad\qquad\qquad 
	\langle \hat{Q}_{sc, \sigma}(t_1)\hat{Q}_{sc, \sigma}(t_2) \hat{a}'_{\bm{k}, \sigma}(0) [\hat{V}(t_4), \hat{Q}_{sc, \sigma}(t_3)] \rangle \nonumber \\
	&= \dfrac{1}{g_0\omega_k\sqrt{\omega_k}} \int_{0}^{t} dt_1  \int_{0}^{t} dt_2 \int_0^t dt_3 e^{-i\omega_k(t_1 + t_2 - t_3)} \int_0^{t_3} dt_4\nonumber \\
	&\qquad\qquad\qquad\qquad 
	\langle \hat{Q}_{sc, \sigma}(t_1)\hat{Q}_{sc, \sigma}(t_2)  [\hat{Q}'_{sc, \lambda}(t_4), \hat{Q}_{sc, \sigma}(t_3)] \rangle \nonumber\\
	&\qquad\qquad\qquad\qquad 
	\cdot\sum_{\bm{q}, \lambda} \dfrac{g_0}{\sqrt{\omega_q}} \langle \hat{a}'_{\bm{k}, \sigma}(0) (\hat{a}_{\bm{q}, \lambda}(0) e^{-i \omega_q t} + \hat{a}_{\bm{q}, \lambda}^\dagger(0) e^{i \omega_q t}) \rangle \nonumber \\
	&= \dfrac{1}{\omega_k^2} \int_{0}^{t} dt_1  \int_{0}^{t} dt_2 \int_0^t dt_3 \int_0^{t_3} dt_4 e^{-i\omega_k(t_1 + t_2 - t_3 - t_4)} \nonumber \\
	&\qquad\qquad\qquad\qquad 
	\langle \hat{Q}_{sc, \sigma}(t_1)\hat{Q}_{sc, \sigma}(t_2)  [\hat{Q}_{sc, \sigma}(t_4), \hat{Q}_{sc, \sigma}(t_3)] \rangle  \\
	\langle \hat{T}^{(2)\dagger}_{\bm{k}, \sigma}(t) \hat{T}^{(0)\dagger}_{\bm{k}, \sigma}(t) \hat{T}^{(0)}_{\bm{k}, \sigma}(t) \hat{T}^{(2)}_{\bm{k}, \sigma}(t) \rangle
	&= -\dfrac{1}{g_0^2\omega_k} \int_{0}^{t} dt_1 \int_0^t dt_3 e^{-i\omega_k(t_1 - t_3)}  \int_{0}^{t_1} dt_2 \int_0^{t_3} dt_4 \nonumber \\
	&\qquad\qquad\qquad\qquad 
	\langle  [\hat{V}(t_2), \hat{Q}_{sc, \sigma}(t_1)] \hat{a}_{\bm{k}, \sigma}'^\dagger (0)\hat{a}'_{\bm{k}, \sigma}(0)  [\hat{V}(t_4), \hat{Q}_{sc, \sigma}(t_3)] \rangle \nonumber \\
	&=-\dfrac{1}{\omega_k} \int_{0}^{t} dt_1 \int_{0}^{t_1} dt_2 \int_0^t dt_3 \int_0^{t_3} dt_4 e^{-i\omega_k(t_1 + t_2 - t_3 - t_4)} \nonumber \\
	&\qquad\qquad\qquad\qquad 
	\langle [\hat{Q}_{sc, \sigma}(t_2)\hat{Q}_{sc, \sigma}(t_1)]  [\hat{Q}_{sc, \sigma}(t_4), \hat{Q}_{sc, \sigma}(t_3)] \rangle ,
\end{align}
where we have used that $\hat{Q}_{sc, \sigma}(t) = \sum_{j=1}^N \hat{p}_{sc,\sigma, j}(t)$ and $\hat{V}(t)$ are Hermitian operators. 
Before writing out the four-operator electronic expectation value, we will utilize the general result $\int_0^t dt_1 \int_0^t dt_2 f(t_1, t_2) = \int_0^t dt_1 \int_0^{t_1} dt_2 f(t_1, t_2) + \int_0^t dt_1 \int_0^{t_1} dt_2 f(t_2, t_1)$ and collect what would otherwise be seven different terms. With this simplification, we are left with
\begin{align}
	\langle \hat{a}_{\bm{k}, \sigma}'^\dagger (t) \hat{a}_{\bm{k}, \sigma}'^\dagger (t) \hat{a}_{\bm{k}, \sigma}' (t) \hat{a}_{\bm{k}, \sigma}' (t)\rangle 
	&\simeq 
	\dfrac{4g_0^4}{\omega_k^2} \int_{0}^{t} dt_1 \int_{0}^{t_1} dt_2 \int_0^t dt_3 \int_0^{t_3} dt_4 
	e^{-i\omega_k(t_1 + t_2 - t_3 - t_4)} \nonumber\\
	&\qquad\qquad\qquad\langle \hat{Q}_{sc, \sigma}(t_2)\hat{Q}_{sc, \sigma}(t_1)\hat{Q}_{sc, \sigma}(t_3)\hat{Q}_{sc, \sigma}(t_4) \rangle.
\end{align}
We now write out the electronic expectation value following the procedure in App. (\ref{App:Exp_electron_operators}), obtaining an expression of the general form
\begin{equation}
	g_{\bm{k}, \sigma}^{(2)}(0) = \dfrac{\dfrac{N!}{(N-4)!} D^{(0)}_{\bm{k}, \sigma} + \dfrac{N!}{(N-3)!} D^{(2)}_{\bm{k}, \sigma} + \dfrac{N!}{(N-2)!} (D^{(2,2)}_{\bm{k}, \sigma} + D^{(3)}_{\bm{k}, \sigma}) + \dfrac{N!}{(N-1)!} D^{(4)}_{\bm{k}, \sigma}}{N^2(N-1)^2 D^{(0)}_{\bm{k}, \sigma} + 2N^2 (N-1) \sqrt{D^{(0)}_{\bm{k}, \sigma}} \sqrt{\tilde{D}^{(2)}_{\bm{k}, \sigma}} + N^2 \tilde{D}^{(2)}_{\bm{k}, \sigma}}. \label{eq:g2_perturbativeform}
\end{equation}
The coefficients are
\begin{subequations} \label{eq:g2_coeffs_perturbative}
\begin{align}
	D^{(0)}_{\bm{k}, \sigma} 
	&= \dfrac{4g_0^4}{\omega_k^2}  \int_{0}^{t} dt_1 \int_{0}^{t_1} dt_2 \int_0^t dt_3 \int_0^{t_3} dt_4  e^{-i\omega_k(t_1 + t_2 - t_3 - t_4)}
	\langle \hat{{p}}_{sc,\sigma}(t_1) \rangle\langle \hat{{p}}_{sc,\sigma}(t_2) \rangle\langle \hat{{p}}_{sc,\sigma}(t_3) \rangle\langle \hat{{p}}_{sc,\sigma}(t_4) \rangle \nonumber \\
	&= \dfrac{g_0^4}{\omega_k^2} \bigg \lvert \int_0^t dt_1 e^{-i \omega_k t_1} \langle \hat{{p}}_{sc,\sigma}(t_1) \rangle \bigg \rvert^4 \label{eq:g2_coeff_D0}\\
	D^{(2)}_{\bm{k}, \sigma} 
	&=\dfrac{4g_0^4}{\omega_k^2} \int_{0}^{t} dt_1 \int_{0}^{t_1} dt_2 \int_0^t dt_3 \int_0^{t_3} dt_4 
	e^{-i\omega_k(t_1 + t_2 - t_3 - t_4)} \nonumber \\
	&\qquad\qquad \bigg (
	\langle \hat{{p}}_{sc,\sigma}(t_2)\hat{{p}}_{sc,\sigma}(t_1) \rangle\langle \hat{{p}}_{sc,\sigma}(t_3) \rangle\langle \hat{{p}}_{sc,\sigma}(t_4) \rangle  
	+\langle \hat{{p}}_{sc,\sigma}(t_2)\hat{{p}}_{sc,\sigma}(t_3) \rangle\langle \hat{{p}}_{sc,\sigma}(t_1) \rangle\langle \hat{{p}}_{sc,\sigma}(t_4) \rangle \nonumber \\
	&\qquad\qquad
	+\langle \hat{{p}}_{sc,\sigma}(t_2)\hat{{p}}_{sc,\sigma}(t_4) \rangle\langle \hat{{p}}_{sc,\sigma}(t_1) \rangle\langle \hat{{p}}_{sc,\sigma}(t_3) \rangle 
	+\langle \hat{{p}}_{sc,\sigma}(t_1)\hat{{p}}_{sc,\sigma}(t_3) \rangle\langle \hat{{p}}_{sc,\sigma}(t_2) \rangle\langle \hat{{p}}_{sc,\sigma}(t_3) \rangle \nonumber \\
	&\qquad\qquad
	+\langle \hat{{p}}_{sc,\sigma}(t_1)\hat{{p}}_{sc,\sigma}(t_4) \rangle\langle \hat{{p}}_{sc,\sigma}(t_2) \rangle\langle \hat{{p}}_{sc,\sigma}(t_3) \rangle 
	+\langle \hat{{p}}_{sc,\sigma}(t_3)\hat{{p}}_{sc,\sigma}(t_4) \rangle\langle \hat{{p}}_{sc,\sigma}(t_2) \rangle\langle \hat{{p}}_{sc,\sigma}(t_1) \rangle \bigg) \nonumber \\
	&= \dfrac{4g_0^4}{\omega_k^2} \bigg( \int_{0}^{t} dt_1 \int_{0}^{t} dt_2 \int_0^t dt_3 \int_0^{t} dt_4 
	e^{-i\omega_k(t_1 + t_2 - t_3 - t_4)} \langle \hat{{p}}_{sc,\sigma}(t_2)\hat{{p}}_{sc,\sigma}(t_3) \rangle\langle \hat{{p}}_{sc,\sigma}(t_1) \rangle\langle \hat{{p}}_{sc,\sigma}(t_4) \rangle \nonumber \\
	&\qquad 
	+2\text{Re} \bigg \{  \int_{0}^{t} dt_1 \int_{0}^{t_1} dt_2 \int_0^t dt_3 \int_0^{t_3} dt_4 e^{-i\omega_k(t_1 + t_2 - t_3 - t_4)}   \langle \hat{{p}}_{sc,\sigma}(t_2)\hat{{p}}_{sc,\sigma}(t_1) \rangle\langle \hat{{p}}_{sc,\sigma}(t_3) \rangle\langle \hat{{p}}_{sc,\sigma}(t_4) \rangle \bigg \} 
	\bigg) \nonumber \\
	&=\dfrac{4g_0^4}{\omega_k^2} \bigg( \int_{0}^{t} dt_2 \int_0^t dt_3 
	e^{-i\omega_k(t_2 - t_3)} \langle \hat{{p}}_{sc,\sigma}(t_2)\hat{{p}}_{sc,\sigma}(t_3) \rangle\cdot  \bigg \lvert \int_0^t dt_1 e^{i \omega_k t_1} \langle  \hat{{p}}_{sc,\sigma}(t_1) \rangle \bigg \rvert^2 \nonumber \\
	&\qquad 
	+\text{Re} \bigg \{  \int_{0}^{t} dt_1 \int_{0}^{t_1} dt_2 e^{-i\omega_k(t_1 + t_2)}   \langle \hat{{p}}_{sc,\sigma}(t_2)\hat{{p}}_{sc,\sigma}(t_1) \rangle \cdot \bigg ( \int_0^t dt_3 e^{i \omega_k t_3} \langle  \hat{{p}}_{sc,\sigma}(t_3) \rangle \bigg )^2 \bigg \} 
	\bigg)  \\
	D^{(3)}_{\bm{k}, \sigma} &= \dfrac{4g_0^4}{\omega_k^2} \int_{0}^{t} dt_1 \int_{0}^{t_1} dt_2 \int_0^t dt_3 \int_0^{t_3} dt_4 
	e^{-i\omega_k(t_1 + t_2 - t_3 - t_4)}\nonumber \\
	&\qquad\qquad \bigg( 
	\langle \hat{{p}}_{sc,\sigma}(t_2)\hat{{p}}_{sc,\sigma}(t_1) \rangle\langle \hat{{p}}_{sc,\sigma}(t_3) \hat{{p}}_{sc,\sigma}(t_4) \rangle  \nonumber \\
	&\qquad\qquad
	+\langle \hat{{p}}_{sc,\sigma}(t_2)\hat{{p}}_{sc,\sigma}(t_3) \rangle\langle \hat{{p}}_{sc,\sigma}(t_1) \hat{{p}}_{sc,\sigma}(t_4) \rangle \nonumber
	+\langle \hat{{p}}_{sc,\sigma}(t_2)\hat{{p}}_{sc,\sigma}(t_4) \rangle\langle \hat{{p}}_{sc,\sigma}(t_1) \hat{{p}}_{sc,\sigma}(t_3) \rangle \nonumber \\		
	&\qquad\qquad
	+\langle \hat{{p}}_{sc,\sigma}(t_2)\hat{{p}}_{sc,\sigma}(t_1) \hat{{p}}_{sc,\sigma}(t_3) \rangle\langle  \hat{{p}}_{sc,\sigma}(t_4) \rangle
	+\langle \hat{{p}}_{sc,\sigma}(t_2)\hat{{p}}_{sc,\sigma}(t_1) \hat{{p}}_{sc,\sigma}(t_4) \rangle\langle  \hat{{p}}_{sc,\sigma}(t_3) \rangle\nonumber \\		
	&\qquad\qquad
	+\langle \hat{{p}}_{sc,\sigma}(t_2)\hat{{p}}_{sc,\sigma}(t_3) \hat{{p}}_{sc,\sigma}(t_4) \rangle\langle  \hat{{p}}_{sc,\sigma}(t_1) \rangle
	+\langle \hat{{p}}_{sc,\sigma}(t_1)\hat{{p}}_{sc,\sigma}(t_3) \hat{{p}}_{sc,\sigma}(t_4) \rangle\langle  \hat{{p}}_{sc,\sigma}(t_2) \rangle
	\bigg) \nonumber \\
	&= \dfrac{2g_0^4}{\omega_k^2} \bigg( 2\bigg \lvert \int_{0}^{t} dt_1 \int_{0}^{t_1} dt_2 e^{-i\omega_k(t_1 + t_2)} \langle \hat{{p}}_{sc,\sigma}(t_2)\hat{{p}}_{sc,\sigma}(t_1) \rangle \bigg \rvert^2 \nonumber \\
	&\qquad\qquad + \bigg ( \int_{0}^{t} dt_1 \int_{0}^{t} dt_3 e^{-i\omega_k(t_1 - t_3)} \langle \hat{{p}}_{sc,\sigma}(t_1)\hat{{p}}_{sc,\sigma}(t_3) \rangle \bigg )^2 \nonumber \\
	&\qquad\qquad + 4\text{Re} \bigg\{\int_{0}^{t} dt_1 \int_{0}^{t_1} dt_2 \int_0^t dt_3 \int_0^{t} dt_4 
	e^{-i\omega_k(t_1 + t_2 - t_3 - t_4)}
	\langle \hat{{p}}_{sc,\sigma}(t_2)\hat{{p}}_{sc,\sigma}(t_1) \hat{{p}}_{sc,\sigma}(t_3) \rangle\langle  \hat{{p}}_{sc,\sigma}(t_4) \rangle
	\bigg\} 
	\bigg) \\
	D^{(4)}_{\bm{k}, \sigma} &= \dfrac{4g_0^4}{\omega_k^2}  \int_{0}^{t} dt_1 \int_{0}^{t_1} dt_2 \int_0^t dt_3 \int_0^{t_3} dt_4  e^{-i\omega_k(t_1 + t_2 - t_3 - t_4)}
	\langle \hat{{p}}_{sc,\sigma}(t_2) \hat{{p}}_{sc,\sigma}(t_1) \hat{{p}}_{sc,\sigma}(t_3) \hat{{p}}_{sc,\sigma}(t_4) \rangle
	\\
	\tilde{D}^{(2)}_{\bm{k}, \sigma} &= \dfrac{g_0^4}{\omega_k^2} \bigg ( \int_{0}^{t} dt_1 \int_{0}^{t} dt_3 e^{-i\omega_k(t_1 - t_2)} \langle \hat{{p}}_{sc,\sigma}(t_1)\hat{{p}}_{sc,\sigma}(t_2) \rangle \bigg )^2, \label{eq:g2_coeff_D2tilde}
\end{align}
\end{subequations}
where the expression for $\tilde{D}^{(2)}_{\bm{k}, \sigma}$ stems from the counting operator in the denominator of $g_{\bm{k}, \sigma}^{(2)}(0)$, which coincides with Eq. (\ref{eq:counting_operator_exact}) when the exact dynamics of the emitter is replaced by the semiclassical form as discussed in the main text. Inserting the expressions for the coefficients in Eq. (\ref{eq:g2_coeffs_perturbative}) into Eq. (\ref{eq:g2_perturbativeform}) concludes the derivation of the expression for $g_{\bm{k}, \sigma}^{(2)}(0)$ in the PHD given in Eq. (\ref{eq:g2_perturbative}) in the main text.

We note that all terms in the denominator of $g_{\bm{k}, \sigma}^{(2)}(0)$ scale as $g_0^4$, yielding a PHD expression for $g_{\bm{k}, \sigma}^{(2)}(0)$, which is independent of $g_0$.

\section{Expectation values of electronic operators} \label{App:Exp_electron_operators}
In this appendix, we show how we calculate expectation values for the electronic operators entering both the expressions for the spectrum [Eq. (\ref{eq:spectrum_perturbative})], quadrature variance [Eq. (\ref{eq:quadrature_variance_peturbative})], and second-order correlation function [Eq. (\ref{eq:g2_perturbative})] in the PHD with equations in the main text and how we obtain their scaling relations with $N$. We outline a general example, as the electronic expectation values over products of $\hat{\bm{e}}_\sigma \cdot \hat{\bm{Q}}(t)$ are expressed in terms of $\hat{{p}}_\sigma(t)$. We show how such an expression in the PHD can be rewritten in terms of transition dipoles, paving the way for numerical evaluation.  
We start with the general expression
\begin{equation}
	\langle \hat{Q}_{\sigma}(t_1) \hat{Q}_{\sigma}(t_2) \hat{Q}_{\sigma}(t_3) \hat{Q}_{\sigma}(t_4) \rangle, \label{eq:Q_prod_exp_val}
\end{equation}
where $\hat{Q}_{\sigma}(t)$ has been introduced as short-hand notation for $\hat{\bm{e}}_\sigma \cdot \hat{\bm{Q}}(t)$. Recalling that $\hat{Q}_{\sigma}(t) = \sum_{j=1}^N \hat{\bm{e}}_\sigma \cdot \hat{\bm{p}}_j(t)$ is a sum over independent emitters, which we take to be identical, we obtain a sum of $N^4$ expectation values over four-fold products of $\hat{\bm{e}}_\sigma \cdot \hat{\bm{p}}_j(t)$,
\begin{equation}
	\sum_{i,j,k,l}^N\langle \hat{\bm{e}}_\sigma \cdot \hat{\bm{p}}_i(t_1) \hat{\bm{e}}_\sigma \cdot \hat{\bm{p}}_j(t)(t_2) \hat{\bm{e}}_\sigma \cdot \hat{\bm{p}}_k(t_3) \hat{\bm{e}}_\sigma \cdot \hat{\bm{p}}_l(t_4) \rangle
\end{equation} 
Such an expectation value will factorize completely in the case $i\neq j \neq k \neq l$, not at all in the case $i = j = k = l$, and partly in the cases when some indeces are identical and others not. Of these we can consider $i=j=k \neq l$, $i = j \neq k=l$ and $i = j \neq k\neq l$, along with all unique permutations of indices, which will result in a binomial coefficient for a front factor. The generalization of Eq. (\ref{eq:independent_and_identical_emitters}) in the case of a four-fold product becomes
\begin{subequations}
\begin{align}
	\langle \hat{Q}_{\sigma}(t_1) \hat{Q}_{\sigma}(t_2) &\hat{Q}_{\sigma}(t_3) \hat{Q}_{\sigma}(t_4) \rangle = \nonumber \\
	& N\langle \hat{{p}}_\sigma(t_1) \hat{{p}}_\sigma(t_2) \hat{{p}}_\sigma(t_3) \hat{{p}}_\sigma(t_4) \rangle   &i = j = k = l \\
	&+\dfrac{N!}{(N-2)!}
	\bigg[ \langle \hat{{p}}_\sigma(t_1) \hat{{p}}_\sigma(t_2) \hat{{p}}_\sigma(t_3) \rangle\langle \hat{{p}}_\sigma(t_4) \rangle
	+\langle \hat{{p}}_\sigma(t_1) \hat{{p}}_\sigma(t_2) \hat{{p}}_\sigma(t_4) \rangle\langle \hat{{p}}_\sigma(t_3) \rangle    \nonumber \\
	&\qquad\qquad \  +\langle \hat{{p}}_\sigma(t_1) \hat{{p}}_\sigma(t_3) \hat{{p}}_\sigma(t_4) \rangle\langle \hat{{p}}_\sigma(t_2) \rangle
	+\langle \hat{{p}}_\sigma(t_2) \hat{{p}}_\sigma(t_3) \hat{{p}}_\sigma(t_4) \rangle\langle \hat{{p}}_\sigma(t_1) \rangle    \bigg] &i=j=k \neq l  \\
	&+\dfrac{N!}{(N-2)!} 
	\bigg[ \langle \hat{{p}}_\sigma(t_1) \hat{{p}}_\sigma(t_2) \rangle\langle \hat{{p}}_\sigma(t_3) \hat{{p}}_\sigma(t_4) \rangle + \langle \hat{{p}}_\sigma(t_1) \hat{{p}}_\sigma(t_3) \rangle \langle \hat{{p}}_\sigma(t_2)  \hat{{p}}_\sigma(t_4) \rangle \nonumber\\
	&\qquad\qquad \  +\langle \hat{{p}}_\sigma(t_1) \hat{{p}}_\sigma(t_4) \rangle \langle \hat{{p}}_\sigma(t_2)  \hat{{p}}_\sigma(t_3) \rangle \bigg] &i = j \neq k=l  \\
	&+ \dfrac{N!}{(N-3)!}
	\bigg[  \langle \hat{{p}}_\sigma(t_1) \hat{{p}}_\sigma(t_2) \rangle\langle \hat{{p}}_\sigma(t_3) \rangle\langle \hat{{p}}_\sigma(t_4) \rangle 
	+ \langle \hat{{p}}_\sigma(t_1) \hat{{p}}_\sigma(t_3) \rangle\langle \hat{{p}}_\sigma(t_2) \rangle\langle \hat{{p}}_\sigma(t_4) \rangle \nonumber \\
	&\qquad\qquad \  +\langle \hat{{p}}_\sigma(t_1) \hat{{p}}_\sigma(t_4) \rangle\langle \hat{{p}}_\sigma(t_2) \rangle\langle \hat{{p}}_\sigma(t_3) \rangle 
	+ \langle \hat{{p}}_\sigma(t_2) \hat{{p}}_\sigma(t_3) \rangle\langle \hat{{p}}_\sigma(t_1) \rangle\langle \hat{{p}}_\sigma(t_4) \rangle \nonumber \\
	&\qquad\qquad \  +\langle \hat{{p}}_\sigma(t_2) \hat{{p}}_\sigma(t_4) \rangle\langle \hat{{p}}_\sigma(t_1) \rangle\langle \hat{{p}}_\sigma(t_3) \rangle 
	+ \langle \hat{{p}}_\sigma(t_3) \hat{{p}}_\sigma(t_4) \rangle\langle \hat{{p}}_\sigma(t_1) \rangle\langle \hat{{p}}_\sigma(t_2) \rangle 
	\bigg] &i = j \neq k\neq l  \label{subeq:el_exp_value_N3}\\
	&+ \dfrac{N!}{(N-4)!} \bigg[ \langle \hat{{p}}_\sigma(t_1) \rangle\langle \hat{{p}}_\sigma(t_2) \rangle\langle \hat{{p}}_\sigma(t_3) \rangle\langle \hat{{p}}_\sigma(t_4) \rangle \bigg]. &i\neq j \neq k \neq l \label{subeq:el_exp_value_N4}
\end{align}
\end{subequations}
In PHD, we work with products of $\hat{Q}_{sc, \sigma}(t) = \sum_{j=1}^N \hat{p}_{sc,\sigma, j}(t)$. The treatment of these is completely equivalent to the one detailed above. 

An electronic expectation value like $\langle \hat{{p}}_\sigma(t_1) \hat{{p}}_\sigma(t_2) \rangle$ can be expressed in terms of transition dipoles, $p_{mi} (t) = \bra{\phi_m} \hat{p}_{sc,\sigma}(t) \ket{\phi_i}$, by inserting complete sets of states for the emitter system,
\begin{align}
	\langle \hat{{p}}_\sigma(t_1) \hat{{p}}_\sigma(t_2) \rangle 
	&= \sum_m \bra{\phi_i} \hat{p}_{sc,\sigma}(t_1) \ket{\phi_m}\bra{\phi_m} \hat{p}_{sc,\sigma}(t_2) \ket{\phi_i} 
	= \sum_m p_{im}(t_1) p_{mi} (t_2) \nonumber
	\intertext{and similarly}
	\langle \hat{{p}}_\sigma(t_1) \hat{{p}}_\sigma(t_2) \hat{{p}}_\sigma(t_3) \rangle 
	&= \sum_{m,n} \bra{\phi_i} \hat{p}_{sc,\sigma}(t_1) \ket{\phi_m}\bra{\phi_m} \hat{p}_{sc,\sigma}(t_2) \ket{\phi_n}\bra{\phi_n}  \hat{{p}}_\sigma(t_3) \ket{\phi_i} = \sum_{m,n} p_{im}(t_1) p_{mn} (t_2) p_{ni} (t_3), \nonumber
\end{align}
which can be evaluated numerically by obtaining all the time-evolved eigenstates of the emitter system. These transition diples are then integrated in accordance with the obtained expressions for observables, letting $t$ be a time after the driving pulse is ended. As discussed in Sec. \ref{Sec:Discussion}, this highlights that the physical origin of these transition dipoles is related to expressions of higher moments of the photonic expectation values, which results in higher moment of emitter expectation values.

\section{Derivation of expression values in the displaced frame} \label{App:expectation_value_expression}
	In this appendix, we show how we obtain Eq. (\ref{eq:expectation_value_displaced}) in the main text. By performing the transformations often used in the Schrödinger picture \cite{Gorlach2020, Lewenstein2021, Lange2024a, Lange2025a, Lange2025b, Yi2024}, we first go to the rotating frame of the free-field Hamiltonian, then displace away the driving field before going into the semiclassical frame. Applying these transformations yields
	\begin{align}
		\ket{\tilde{\Psi}(t)} &= \hat{\mathcal{U}}_{sc}^\dagger(t) \hat{D}^\dagger(\alpha) \mathcal{U}_F^\dagger(t) \ket{\Psi(t)} \nonumber \\
		& = \hat{\mathcal{U}}_{sc}^\dagger(t) \hat{D}^\dagger(\alpha) \mathcal{U}_F^\dagger(t) \hat{\mathcal{U}}(t) \ket{\Psi(0)},
	\end{align} 
	where we denote $\alpha = \alpha_{\bm{k}_L, \sigma_L}$ for convenience. The state $\ket{\tilde{\Psi}(t)}$ satisfies the TDSE
	\begin{equation}
		i \partial_t 	\ket{\tilde{\Psi}(t)} = \hat{V}(t) 	\ket{\tilde{\Psi}(t)}, \label{eq:app_transformed_TDSE}
	\end{equation}
	with $\hat{V}(t)$ given in Eq. (\ref{eq:V_definition}). As similar transformations must be made on the operators, we obtain the following expression for a general operator $\hat{O}(t)$
	\begin{equation}
		\langle \hat{O}(t) \rangle = \bra{\Psi(0)} \mathcal{U}^\dagger (t) \hat{U}_{F}(t) \hat{D}(\alpha) \hat{U}_{sc}(t) \hat{U}^\dagger_{sc}(t) \hat{D}^\dagger(\alpha) \hat{\mathcal{U}}^\dagger_F(t) \hat{O}_S(t) \hat{\mathcal{U}}_F(t) \hat{D}(\alpha) \hat{\mathcal{U}}_{sc}(t) \hat{\mathcal{U}}_{sc}^\dagger (t) \hat{D}^\dagger(\alpha) \hat{\mathcal{U}}_F^\dagger(t) \hat{\mathcal{U}}(t) \ket{\Psi(0)}, \label{eq:app_exp_val1}
	\end{equation}
	where $\hat{\mathcal{U}}(t)$ is the exact time-evolution operator for the full system, and the operator $\hat{O}_S(t)$ denotes the (possibly time-dependent) operator $\hat{O}$ in the Schrödinger picture. We now insert $\hat{D}(\alpha) \hat{D}^\dagger(\alpha) = \mathbb{1}$ into Eq. (\ref{eq:app_exp_val1}) and obtain
	\begin{equation}
		\langle \hat{O}(t) \rangle = \bra{\tilde{\Psi}(0)} \hat{\mathcal{U}}'^\dagger (t) \hat{O}'_S(t) \hat{\mathcal{U}}' (t) \ket{\tilde{\Psi}(0)}, \label{eq:app_exp_val2}
	\end{equation}
	where we have defined
	\begin{align}
		\ket{\tilde{\Psi}(0)} &= \hat{D}^\dagger(\alpha) \ket{\Psi(0)}, \\
		\hat{\mathcal{U}}'(t)&= \hat{D}^\dagger(\alpha) \hat{\mathcal{U}}^\dagger_{sc}(t) \hat{\mathcal{U}}_F^\dagger(t) \hat{\mathcal{U}}(t) \hat{D}(\alpha), \label{eq:app_exp_val3} \\
		\hat{O}'_S(t) &= \hat{\mathcal{U}}^\dagger_{sc}(t) \hat{D}^\dagger(\alpha) \hat{\mathcal{U}}_{F}^\dagger(t) \hat{O}_S(t) \hat{\mathcal{U}}_F(t) \hat{D}(\alpha) \hat{\mathcal{U}}_{sc}(t),\label{eq:app_exp_val4}
	\end{align}
	as the transformed initial state, time-evolution operator, and operator whose expectation value we want to calculate, respectively, which are all given in the main text. 
	
	From Eq. (\ref{eq:app_transformed_TDSE}), it is seen that the time-evolution operator $	\hat{\mathcal{U}}'(t)$ satisfies the TDSE
	\begin{equation}
		i \partial_t ~ 	\hat{\mathcal{U}}'(t) = \hat{V}(t) 	\hat{\mathcal{U}}'(t).
	\end{equation}
	
	By taking $\hat{O}_S(t) = \hat{a}_{\bm{k}, \sigma}$, one can use the transformations given in Eq. (\ref{eq:app_exp_val2})-(\ref{eq:app_exp_val4}) and obtain an Heisenberg equation of motion for the transformed operator, $\hat{a}_{\bm{k}, \sigma}'(t)$, the solution to which is given in Eq. (\ref{eq:Heisenberg_transformed_exact}).
	
\section{Higher-order contributions to the counting operator} \label{App:higher_order_corrections}
In this appendix, we calculate the higher-order contribution to the PHD expression for the harmonic spectrum given in Eq. (\ref{eq:spectrum_perturbative}) in the main text. We show that the leading order correction is on order $\mathcal{O}(g_0^2 N^3)$ which is not compliant with the selection rules obeyed by the coherent contribution also seen in experimental results.

To evaluate the spectrum of emitted harmonics, we must consider the counting operator $\langle \hat{a}_{\bm{k}, \sigma}^\dagger \hat{a}_{\bm{k}, \sigma} \rangle$, which we can expand in powers of $g_0$ in the PHD using Eq. (\ref{eq:Heisenberg_perturbative}). Consulting Eq. (\ref{eq:terms_of_a_perturbative}), we find that the only contribution up to order $g_0^2$ stems from $\langle \hat{T}^{(1)}_{\bm{k}, \sigma}(t) \hat{T}^{(1)}_{\bm{k}, \sigma}(t) \rangle$, which is the only $\mathcal{O}(g_0^2)$ term that does not vanish when calculating the expectation value. This results in the terms $\sqrt{D^{(0)}_{\bm{k}, \sigma}}$ and $\sqrt{\tilde{D}^{(2)}_{\bm{k}, \sigma}}$ of Eqs. (\ref{eq:g2_coeff_D0}) and (\ref{eq:g2_coeff_D2tilde}) above. As discussed in the main text, this lowest order expansion sees no effects of quantized field fluctuations and shows complete decoupling of harmonic modes. 

Going to higher order in $g_0 N$, we immediately find that all $\mathcal{O}(g_0^3)$ contributions vanish, since all $\mathcal{O}(g_0^3)$ combinations of the operators in Eq. (\ref{eq:terms_of_a_perturbative}) will have an odd number of photonic operators. 

For $\mathcal{O}(g_0^4)$, we find several nonvanishing terms that consitute a correction to the counting operator, 
\begin{align}
	\langle \hat{a}_{\bm{k}, \sigma}'^\dagger (t) \hat{a}_{\bm{k}, \sigma}' (t) \rangle_{cor} 
	&= 	
	g_0^4 \bigg( \langle \hat{T}^{(2)\dagger}_{\bm{k}, \sigma}(t) \hat{T}^{(2)}_{\bm{k}, \sigma}(t) \rangle  
	+ \langle \hat{T}^{(1)\dagger}_{\bm{k}, \sigma}(t) \hat{T}^{(3a)}_{\bm{k}, \sigma}(t) \rangle 
	+ \langle \hat{T}^{(3a)\dagger}_{\bm{k}, \sigma}(t) \hat{T}^{(1)}_{\bm{k}, \sigma}(t) \rangle \nonumber\\
	&\qquad\qquad\qquad+\langle \hat{T}^{(1)\dagger}_{\bm{k}, \sigma}(t) \hat{T}^{(3b)}_{\bm{k}, \sigma}(t) \rangle 
	+\langle \hat{T}^{(3b)\dagger}_{\bm{k}, \sigma}(t) \hat{T}^{(1)}_{\bm{k}, \sigma}(t) \rangle \bigg),
\end{align}
where we can evaluate the photonic operators, $\langle \hat{a}_{\bm{q}, \lambda}(0) \hat{a}_{\bm{p}, \rho}^\dagger(0)\rangle = \delta_{\bm{q},\bm{p}} \delta_{\lambda, \rho}$, obtaining
\begin{subequations}
\begin{align}
	\langle \hat{T}^{(2)\dagger}_{\bm{k}, \sigma}(t) \hat{T}^{(2)}_{\bm{k}, \sigma}(t) \rangle 
	&=
	\dfrac{-1}{g_0^2 \omega_k}  \int_{0}^{t} dt_1  \int_{0}^{t_1} dt_2 \int_{0}^{t} dt_3  \int_{0}^{t_3} dt_4 e^{-i\omega_k(t_1 - t_3)} \nonumber \\
	&\qquad\qquad\qquad
	\langle [\hat{Q}_{sc, \sigma}(t_1)\hat{Q}_{sc, \lambda}(t_2)][\hat{Q}_{sc, \sigma}(t_3)\hat{Q}_{sc, \rho}(t_4)] \rangle  \nonumber \\
	&\qquad\qquad\qquad
	\cdot\sum_{\bm{q},\bm{p}, \lambda,\rho} \dfrac{g_0^2}{\sqrt{\omega_q \omega_p}} \langle 
	(\hat{a}_{\bm{q}, \lambda}(0) e^{-i \omega_q t_2} + \hat{a}_{\bm{q}, \lambda}^\dagger(0) e^{i \omega_q t_2})
	(\hat{a}_{\bm{p}, \rho}(0) e^{-i \omega_p t_4} + \hat{a}_{\bm{p}, \rho}^\dagger(0) e^{i \omega_p t_4}) \rangle \nonumber \\
	&=
	-\sum_{\bm{q}, \lambda}
	\dfrac{1}{\omega_k \omega_q}  \int_{0}^{t} dt_1  \int_{0}^{t_1} dt_2 \int_{0}^{t} dt_3  \int_{0}^{t_3} dt_4 e^{-i\omega_k(t_1 - t_3) -i\omega_q(t_2 - t_4)} \nonumber \\
	&\qquad\qquad\qquad
	\langle [\hat{Q}_{sc, \sigma}(t_1)\hat{Q}_{sc, \lambda}(t_2)][\hat{Q}_{sc, \sigma}(t_3)\hat{Q}_{sc, \lambda}(t_4)] \rangle \label{subeq:spec_higher_order_terms_1} \\
	\langle \hat{T}^{(1)\dagger}_{\bm{k}, \sigma}(t) \hat{T}^{(3a)}_{\bm{k}, \sigma}(t) \rangle 
	&= 
	\dfrac{1}{g_0^2 \omega_k}  \int_{0}^{t} dt_1  \int_{0}^{t} dt_2 \int_{0}^{t_2} dt_3  \int_{0}^{t_2} dt_4 e^{-i\omega_k(t_1 - t_2)} \nonumber \\
	&\qquad\qquad\qquad
	\langle \hat{Q}_{sc, \sigma}(t_1)\hat{Q}_{sc, \lambda}(t_3)\hat{Q}_{sc, \sigma}(t_2)\hat{Q}_{sc, \rho}(t_4) \rangle  \nonumber \\
	&\qquad\qquad\qquad
	\cdot\sum_{\bm{q},\bm{p}, \lambda,\rho} \dfrac{g_0^2}{\sqrt{\omega_q \omega_p}} \langle 
	(\hat{a}_{\bm{q}, \lambda}(0) e^{-i \omega_q t_3} + \hat{a}_{\bm{q}, \lambda}^\dagger(0) e^{i \omega_q t_3})
	(\hat{a}_{\bm{p}, \rho}(0) e^{-i \omega_p t_4} + \hat{a}_{\bm{p}, \rho}^\dagger(0) e^{i \omega_p t_4}) \rangle \nonumber \\
	&=\sum_{\bm{q}, \lambda} \dfrac{1}{\omega_k \omega_q}  \int_{0}^{t} dt_1  \int_{0}^{t} dt_2 \int_{0}^{t_2} dt_3  \int_{0}^{t_2} dt_4 e^{-i\omega_k(t_1 - t_2) -i\omega_q(t_3 - t_4)} \nonumber \\
	&\qquad\qquad\qquad
	\langle \hat{Q}_{sc, \sigma}(t_1)\hat{Q}_{sc, \lambda}(t_3)\hat{Q}_{sc, \sigma}(t_2)\hat{Q}_{sc, \lambda}(t_4) \rangle  \label{subeq:spec_higher_order_terms_2}  \\
	\langle \hat{T}^{(3a)\dagger}_{\bm{k}, \sigma}(t) \hat{T}^{(1)}_{\bm{k}, \sigma}(t) \rangle 
	&= \bigg[ \langle \hat{T}^{(1)\dagger}_{\bm{k}, \sigma}(t) \hat{T}^{(3a)}_{\bm{k}, \sigma}(t) \rangle \bigg]^*  \\
	\langle \hat{T}^{(1)\dagger}_{\bm{k}, \sigma}(t) \hat{T}^{(3b)}_{\bm{k}, \sigma}(t) \rangle 
	&= 
	\dfrac{-1}{g_0^2 \omega_k}  \int_{0}^{t} dt_1  \int_{0}^{t} dt_2 \int_{0}^{t_2} dt_3  \int_{0}^{t_3} dt_4 e^{-i\omega_k(t_1 - t_2)} \nonumber \\
	&\qquad\qquad \bigg[
	\langle \hat{Q}_{sc, \sigma}(t_1) \hat{Q}_{sc, \sigma}(t_2) \hat{Q}_{sc, \lambda}(t_3)\hat{Q}_{sc, \rho}(t_4) \rangle  \nonumber \\
	&\qquad\qquad\qquad
	\cdot\sum_{\bm{q},\bm{p}, \lambda,\rho} \dfrac{g_0^2}{\sqrt{\omega_q \omega_p}} \langle 
	(\hat{a}_{\bm{q}, \lambda}(0) e^{-i \omega_q t_3} + \hat{a}_{\bm{q}, \lambda}^\dagger(0) e^{i \omega_q t_3})
	(\hat{a}_{\bm{p}, \rho}(0) e^{-i \omega_p t_4} + \hat{a}_{\bm{p}, \rho}^\dagger(0) e^{i \omega_p t_4}) \rangle \nonumber \\
	&\qquad\qquad
	+\langle \hat{Q}_{sc, \sigma}(t_1) \hat{Q}'_{sc, \rho}(t_4) \hat{Q}_{sc, \lambda}(t_3) \hat{Q}_{sc, \sigma}(t_2) \rangle  \nonumber \\
	&\qquad\qquad\qquad
	\cdot\sum_{\bm{q},\bm{p}, \lambda,\rho} \dfrac{g_0^2}{\sqrt{\omega_q \omega_p}} \langle 
	(\hat{a}_{\bm{p}, \rho}(0) e^{-i \omega_p t_4} + \hat{a}_{\bm{p}, \rho}^\dagger(0) e^{i \omega_p t_4})
	(\hat{a}_{\bm{q}, \lambda}(0) e^{-i \omega_q t_3} + \hat{a}_{\bm{q}, \lambda}^\dagger(0) e^{i \omega_q t_3})
	\rangle \nonumber \bigg] \\
	&=-\sum_{\bm{q}, \lambda} \dfrac{1}{\omega_k\omega_q}  \int_{0}^{t} dt_1  \int_{0}^{t} dt_2 \int_{0}^{t_2} dt_3  \int_{0}^{t_3} dt_4 e^{-i\omega_k(t_1 - t_2)}  \nonumber \\
	&\qquad\qquad\qquad
	\bigg[ e^{ -i\omega_q(t_3 - t_4)} \langle \hat{Q}_{sc, \sigma}(t_1) \hat{Q}_{sc, \sigma}(t_2) \hat{Q}_{sc, \lambda}(t_3)\hat{Q}_{sc, \rho}(t_4)  \rangle  \nonumber \\
	&\qquad\qquad\qquad+ e^{ -i\omega_q(t_4 - t_3)} \langle \hat{Q}_{sc, \sigma}(t_1) \hat{Q}_{sc, \rho}(t_4) \hat{Q}_{sc, \lambda}(t_3) \hat{Q}_{sc, \sigma}(t_2) \rangle \bigg]  \label{subeq:spec_higher_order_terms_3}  \\
	\langle \hat{T}^{(3b)\dagger}_{\bm{k}, \sigma}(t) \hat{T}^{(1)}_{\bm{k}, \sigma}(t) \rangle 
	&= \bigg[ \langle \hat{T}^{(1)\dagger}_{\bm{k}, \sigma}(t) \hat{T}^{(3b)}_{\bm{k}, \sigma}(t) \rangle \bigg]^* .
\end{align}
\label{eq:spec_higher_order_terms}
\end{subequations}
We note that each term now includes a sum over harmonic modes. 

For each term, the electronic expectation value can now be written out according to App. \ref{App:Exp_electron_operators}. The leading order contribution to $\langle \hat{a}_{\bm{k}, \sigma}'^\dagger (t) \hat{a}_{\bm{k}, \sigma}' (t) \rangle_{cor}$ at $\mathcal{O}(g_0^4N^4)$ vanishes, as seen by writing out Eq. (\ref{subeq:el_exp_value_N4}) for each of the terms in Eq. (\ref{eq:spec_higher_order_terms}), 
\begin{subequations}
\begin{align}
	\langle \hat{T}^{(2)\dagger}_{\bm{k}, \sigma}(t) \hat{T}^{(2)}_{\bm{k}, \sigma}(t) \rangle 
	&\simeq \mathcal{O}(N^2)  \\
	\langle \hat{T}^{(1)\dagger}_{\bm{k}, \sigma}(t) \hat{T}^{(3a)}_{\bm{k}, \sigma}(t) \rangle 
	&\simeq N^4 \dfrac{1}{\omega_k}  \int_{0}^{t} dt_1  \int_{0}^{t} dt_2 e^{-i\omega_k(t_1 - t_2)} \langle \hat{p}_{sc, \sigma}(t_1) \rangle \langle \hat{p}_{sc, \sigma}(t_2) \rangle \nonumber \\
	&\qquad\qquad \cdot \sum_{\bm{q}, \lambda} \dfrac{1}{\omega_q}
	\int_{0}^{t_2} dt_3  \int_{0}^{t_2} dt_4 e^{ -i\omega_q(t_3 - t_4)} 
	\langle \hat{p}_{sc, \lambda}(t_3) \rangle \langle \hat{p}_{sc, \lambda}(t_4) \rangle + \mathcal{O}(N^3)  \\
	\langle \hat{T}^{(1)\dagger}_{\bm{k}, \sigma}(t) \hat{T}^{(3b)}_{\bm{k}, \sigma}(t) \rangle 
	&\simeq -N^4 \dfrac{1}{\omega_k}  \int_{0}^{t} dt_1  \int_{0}^{t} dt_2 e^{-i\omega_k(t_1 - t_2)} \langle \hat{p}_{sc, \sigma}(t_1) \rangle \langle \hat{p}_{sc, \sigma}(t_2) \rangle \nonumber \\
	&\qquad\qquad \cdot \sum_{\bm{q}, \lambda} \dfrac{1}{\omega_q}
	\bigg[ \int_{0}^{t_2} dt_3  \int_{0}^{t_3} dt_4 e^{ -i\omega_q(t_3 - t_4)} 
	\langle \hat{p}_{sc, \lambda}(t_3) \rangle \langle \hat{p}_{sc, \lambda}(t_4) \rangle \nonumber \\
	&\qquad\qquad\qquad\qquad
	+\int_{0}^{t_2} dt_3  \int_{0}^{t_3} dt_4 e^{ -i\omega_q(t_4 - t_3)} 
	\langle \hat{p}_{sc, \lambda}(t_3) \rangle \langle \hat{p}_{sc, \lambda}(t_4) \rangle 
	\bigg] 
	+ \mathcal{O}(N^3) \nonumber \\
	&= -N^4 \dfrac{1}{\omega_k}  \int_{0}^{t} dt_1  \int_{0}^{t} dt_2 e^{-i\omega_k(t_1 - t_2)} \langle \hat{p}_{sc, \sigma}(t_1) \rangle \langle \hat{p}_{sc, \sigma}(t_2) \rangle \nonumber \\
	&\qquad\qquad \cdot \sum_{\bm{q}, \lambda} \dfrac{1}{\omega_q}
	\int_{0}^{t_2} dt_3  \int_{0}^{t_2} dt_4 e^{ -i\omega_q(t_3 - t_4)} 
	\langle \hat{p}_{sc, \lambda}(t_3) \rangle \langle \hat{p}_{sc, \lambda}(t_4) \rangle + \mathcal{O}(N^3) \nonumber \\
	&\simeq -\langle \hat{T}^{(1)\dagger}_{\bm{k}, \sigma}(t) \hat{T}^{(3a)}_{\bm{k}, \sigma}(t) \rangle + \mathcal{O}(N^3),
\end{align}
\end{subequations}
where it has been used that the contribution from Eq. (\ref{subeq:el_exp_value_N4}) cancels for Eq. (\ref{subeq:spec_higher_order_terms_1}) due to the alternating signs from the commutators between electronic operators. Since this consideration also applies to Eq. (\ref{subeq:el_exp_value_N3}), the $\mathcal{O}(N^3)$ contribution similarly vanish for Eq. (\ref{subeq:spec_higher_order_terms_1}). In total we see 
\begin{equation}
	\langle \hat{a}_{\bm{k}, \sigma}'^\dagger (t) \hat{a}_{\bm{k}, \sigma}' (t) \rangle_{cor} = \mathcal{O}(g_0^4N^3).
\end{equation}
which means that the leading order correction to Eq. (\ref{eq:spectrum_perturbative}) scales as $\mathcal{O}(g_0^2 N^3)$ at most. 
Having concluded that the contribution to Eq. (\ref{eq:spec_higher_order_terms}) from Eq. (\ref{subeq:el_exp_value_N4}) vanishes, we turn to Eq. (\ref{subeq:el_exp_value_N3}) to find the leading order correction to the counting operator at $\mathcal{O}(g_0^4N^3)$. As mentioned before, this contribution to Eq. (\ref{subeq:spec_higher_order_terms_1}) vanishes, while the $\mathcal{O}(g_0^4 N^3)$ contributions to Eq. (\ref{subeq:spec_higher_order_terms_2}), $\langle \hat{T}^{(1)\dagger}_{\bm{k}, \sigma}(t) \hat{T}^{(3a)}_{\bm{k}, \sigma}(t) \rangle_{N^3}$, and to Eq. (\ref{subeq:spec_higher_order_terms_3}), $\langle \hat{T}^{(1)\dagger}_{\bm{k}, \sigma}(t) \hat{T}^{(3b)}_{\bm{k}, \sigma}(t) \rangle_{N^3}$, are
\begin{subequations}
\begin{align}
	\langle \hat{T}^{(1)\dagger}_{\bm{k}, \sigma}(t) \hat{T}^{(3a)}_{\bm{k}, \sigma}(t) \rangle_{N^3} 
	&=\sum_{\bm{q}, \lambda} \dfrac{1}{\omega_k \omega_q}  \int_{0}^{t} dt_1  \int_{0}^{t} dt_2 \int_{0}^{t_2} dt_3  \int_{0}^{t_2} dt_4 e^{-i\omega_k(t_1 - t_2) -i\omega_q(t_3 - t_4)} \nonumber \\
	&\qquad\qquad
	\bigg[ \langle \hat{p}_{sc, \sigma}(t_1)\hat{p}_{sc, \lambda}(t_3) \rangle \langle \hat{p}_{sc, \sigma}(t_2)\rangle \langle \hat{p}_{sc, \lambda}(t_4) \rangle+
	\langle \hat{p}_{sc, \sigma}(t_1)\hat{p}_{sc, \sigma}(t_2)\rangle \langle \hat{p}_{sc, \lambda}(t_3)\rangle \langle \hat{p}_{sc, \lambda}(t_4) \rangle
	\nonumber \\
	&\qquad\qquad
	+\langle \hat{p}_{sc, \sigma}(t_2)\hat{p}_{sc, \lambda}(t_4)\rangle \langle \hat{p}_{sc, \sigma}(t_1)\rangle \langle \hat{p}_{sc, \lambda}(t_3) \rangle
	+\langle \hat{p}_{sc, \lambda}(t_3)\hat{p}_{sc, \sigma}(t_2)\rangle \langle \hat{p}_{sc, \sigma}(t_1)\rangle \langle \hat{p}_{sc, \lambda}(t_4) \rangle
	\nonumber \\
	&\qquad\qquad
	+\langle \hat{p}_{sc, \lambda}(t_3)\hat{p}_{sc, \lambda}(t_4)\rangle \langle \hat{p}_{sc, \sigma}(t_2)\rangle \langle \hat{p}_{sc, \sigma}(t_1) \rangle
	+\langle \hat{p}_{sc, \sigma}(t_1)\hat{p}_{sc, \lambda}(t_4)\rangle \langle \hat{p}_{sc, \sigma}(t_2)\rangle \langle \hat{p}_{sc, \lambda}(t_3) \rangle \bigg] \\
	\langle \hat{T}^{(1)\dagger}_{\bm{k}, \sigma}(t) \hat{T}^{(3b)}_{\bm{k}, \sigma}(t) \rangle_{N^3} 
	&=-\sum_{\bm{q}, \lambda} \dfrac{1}{\omega_k\omega_q}  \int_{0}^{t} dt_1  \int_{0}^{t} dt_2 \int_{0}^{t_2} dt_3  \int_{0}^{t_3} dt_4 e^{-i\omega_k(t_1 - t_2)}  \nonumber \\
	&\qquad \bigg\{e^{ -i\omega_q(t_3 - t_4)} \cdot \nonumber\\
	&\qquad\qquad\bigg[ \langle \hat{p}_{sc, \sigma}(t_1)\hat{p}_{sc, \lambda}(t_3) \rangle \langle \hat{p}_{sc, \sigma}(t_2)\rangle \langle \hat{p}_{sc, \lambda}(t_4) \rangle+
	\langle \hat{p}_{sc, \sigma}(t_1)\hat{p}_{sc, \sigma}(t_2)\rangle \langle \hat{p}_{sc, \lambda}(t_3)\rangle \langle \hat{p}_{sc, \lambda}(t_4) \rangle
	\nonumber \\
	&\qquad\qquad
	+\langle \hat{p}_{sc, \sigma}(t_2)\hat{p}_{sc, \lambda}(t_4)\rangle \langle \hat{p}_{sc, \sigma}(t_1)\rangle \langle \hat{p}_{sc, \lambda}(t_3) \rangle
	+\langle \hat{p}_{sc, \sigma}(t_2) \hat{p}_{sc, \lambda}(t_3)\rangle \langle \hat{p}_{sc, \sigma}(t_1)\rangle \langle \hat{p}_{sc, \lambda}(t_4) \rangle
	\nonumber \\
	&\qquad\qquad
	+\langle \hat{p}_{sc, \lambda}(t_3)\hat{p}_{sc, \lambda}(t_4)\rangle \langle \hat{p}_{sc, \sigma}(t_2)\rangle \langle \hat{p}_{sc, \sigma}(t_1) \rangle
	+\langle \hat{p}_{sc, \sigma}(t_1)\hat{p}_{sc, \lambda}(t_4)\rangle \langle \hat{p}_{sc, \sigma}(t_2)\rangle \langle \hat{p}_{sc, \lambda}(t_3) \rangle \bigg] \nonumber\\
	&\qquad +e^{ -i\omega_q(t_4 - t_3)} \cdot \nonumber\\
	&\qquad\qquad\bigg[ \langle \hat{p}_{sc, \sigma}(t_1)\hat{p}_{sc, \lambda}(t_3) \rangle \langle \hat{p}_{sc, \sigma}(t_2)\rangle \langle \hat{p}_{sc, \lambda}(t_4) \rangle+
	\langle \hat{p}_{sc, \sigma}(t_1)\hat{p}_{sc, \sigma}(t_2)\rangle \langle \hat{p}_{sc, \lambda}(t_3)\rangle \langle \hat{p}_{sc, \lambda}(t_4) \rangle
	\nonumber \\
	&\qquad\qquad
	+\langle \hat{p}_{sc, \lambda}(t_4)\hat{p}_{sc, \sigma}(t_2)\rangle \langle \hat{p}_{sc, \sigma}(t_1)\rangle \langle \hat{p}_{sc, \lambda}(t_3) \rangle
	+\langle \hat{p}_{sc, \lambda}(t_3) \hat{p}_{sc, \sigma}(t_2) \rangle \langle \hat{p}_{sc, \sigma}(t_1)\rangle \langle \hat{p}_{sc, \lambda}(t_4) \rangle
	\nonumber \\
	&\qquad\qquad
	+\langle \hat{p}_{sc, \lambda}(t_4) \hat{p}_{sc, \lambda}(t_3)\rangle \langle \hat{p}_{sc, \sigma}(t_2)\rangle \langle \hat{p}_{sc, \sigma}(t_1) \rangle
	+\langle \hat{p}_{sc, \sigma}(t_1)\hat{p}_{sc, \lambda}(t_4)\rangle \langle \hat{p}_{sc, \sigma}(t_2)\rangle \langle \hat{p}_{sc, \lambda}(t_3) \rangle \bigg] \bigg\} \nonumber\\	
	&=-\sum_{\bm{q}, \lambda} \dfrac{1}{\omega_k\omega_q}  \int_{0}^{t} dt_1  \int_{0}^{t} dt_2 \int_{0}^{t_2} dt_3  \int_{0}^{t_3} dt_4 e^{-i\omega_k(t_1 - t_2)}  \nonumber \\
	&\qquad \bigg\{e^{ -i\omega_q(t_3 - t_4)} \cdot \nonumber\\
	&\qquad\qquad\bigg[ 
	\langle \hat{p}_{sc, \sigma}(t_2)\hat{p}_{sc, \lambda}(t_4)\rangle \langle \hat{p}_{sc, \sigma}(t_1)\rangle \langle \hat{p}_{sc, \lambda}(t_3) \rangle
	+\langle \hat{p}_{sc, \sigma}(t_2) \hat{p}_{sc, \lambda}(t_3)\rangle \langle \hat{p}_{sc, \sigma}(t_1)\rangle \langle \hat{p}_{sc, \lambda}(t_4) \rangle \bigg] \nonumber\\
	&\qquad +e^{ -i\omega_q(t_4 - t_3)} \cdot \nonumber\\
	&\qquad\qquad\bigg[ 
	\langle \hat{p}_{sc, \lambda}(t_4)\hat{p}_{sc, \sigma}(t_2)\rangle \langle \hat{p}_{sc, \sigma}(t_1)\rangle \langle \hat{p}_{sc, \lambda}(t_3) \rangle
	+\langle \hat{p}_{sc, \lambda}(t_3) \hat{p}_{sc, \sigma}(t_2) \rangle \langle \hat{p}_{sc, \sigma}(t_1)\rangle \langle \hat{p}_{sc, \lambda}(t_4) \rangle\bigg] \bigg\} \nonumber\\	
	&\quad -\sum_{\bm{q}, \lambda} \dfrac{1}{\omega_k \omega_q}  \int_{0}^{t} dt_1  \int_{0}^{t} dt_2 \int_{0}^{t_2} dt_3  \int_{0}^{t_2} dt_4 e^{-i\omega_k(t_1 - t_2) -i\omega_q(t_3 - t_4)} \nonumber \\
	&\qquad\qquad
	\bigg[ \langle \hat{p}_{sc, \sigma}(t_1)\hat{p}_{sc, \lambda}(t_3) \rangle \langle \hat{p}_{sc, \sigma}(t_2)\rangle \langle \hat{p}_{sc, \lambda}(t_4) \rangle+
	\langle \hat{p}_{sc, \sigma}(t_1)\hat{p}_{sc, \sigma}(t_2)\rangle \langle \hat{p}_{sc, \lambda}(t_3)\rangle \langle \hat{p}_{sc, \lambda}(t_4) \rangle
	\nonumber \\
	&\qquad\qquad
	+\langle \hat{p}_{sc, \lambda}(t_3)\hat{p}_{sc, \lambda}(t_4)\rangle \langle \hat{p}_{sc, \sigma}(t_2)\rangle \langle \hat{p}_{sc, \sigma}(t_1) \rangle
	+\langle \hat{p}_{sc, \sigma}(t_1)\hat{p}_{sc, \lambda}(t_4)\rangle \langle \hat{p}_{sc, \sigma}(t_2)\rangle \langle \hat{p}_{sc, \lambda}(t_3) \rangle \bigg].
	\end{align}
\end{subequations}
Adding the two contributions gives
\begin{align}
	\langle \hat{T}^{(1)\dagger}_{\bm{k}, \sigma}(t) &\hat{T}^{(3a)}_{\bm{k}, \sigma}(t) \rangle_{N^3} + \langle \hat{T}^{(1)\dagger}_{\bm{k}, \sigma}(t) \hat{T}^{(3b)}_{\bm{k}, \sigma}(t) \rangle_{N^3} \nonumber \\
	&=\sum_{\bm{q}, \lambda} \dfrac{1}{\omega_k \omega_q}  \int_{0}^{t} dt_1  \int_{0}^{t} dt_2 \int_{0}^{t_2} dt_3  \int_{0}^{t_2} dt_4 e^{-i\omega_k(t_1 - t_2) -i\omega_q(t_3 - t_4)} \nonumber \\
	&\qquad\qquad
	\bigg[\langle \hat{p}_{sc, \sigma}(t_2)\hat{p}_{sc, \lambda}(t_4)\rangle \langle \hat{p}_{sc, \sigma}(t_1)\rangle \langle \hat{p}_{sc, \lambda}(t_3) \rangle
	+\langle \hat{p}_{sc, \lambda}(t_3)\hat{p}_{sc, \sigma}(t_2)\rangle \langle \hat{p}_{sc, \sigma}(t_1)\rangle \langle \hat{p}_{sc, \lambda}(t_4) \rangle\bigg] \nonumber \\
	&\quad-\sum_{\bm{q}, \lambda} \dfrac{1}{\omega_k\omega_q}  \int_{0}^{t} dt_1  \int_{0}^{t} dt_2 \int_{0}^{t_2} dt_3  \int_{0}^{t_3} dt_4 e^{-i\omega_k(t_1 - t_2)}  \nonumber \\
	&\qquad \bigg\{e^{ -i\omega_q(t_3 - t_4)} \cdot \nonumber\\
	&\qquad\qquad\bigg[ 
	\langle \hat{p}_{sc, \sigma}(t_2)\hat{p}_{sc, \lambda}(t_4)\rangle \langle \hat{p}_{sc, \sigma}(t_1)\rangle \langle \hat{p}_{sc, \lambda}(t_3) \rangle
	+\langle \hat{p}_{sc, \sigma}(t_2) \hat{p}_{sc, \lambda}(t_3)\rangle \langle \hat{p}_{sc, \sigma}(t_1)\rangle \langle \hat{p}_{sc, \lambda}(t_4) \rangle \bigg] \nonumber\\
	&\qquad +e^{ -i\omega_q(t_4 - t_3)} \cdot \nonumber\\
	&\qquad\qquad\bigg[ 
	\langle \hat{p}_{sc, \lambda}(t_4)\hat{p}_{sc, \sigma}(t_2)\rangle \langle \hat{p}_{sc, \sigma}(t_1)\rangle \langle \hat{p}_{sc, \lambda}(t_3) \rangle
	+\langle \hat{p}_{sc, \lambda}(t_3) \hat{p}_{sc, \sigma}(t_2) \rangle \langle \hat{p}_{sc, \sigma}(t_1)\rangle \langle \hat{p}_{sc, \lambda}(t_4) \rangle\bigg] \bigg\} \nonumber\\	
	&=\sum_{\bm{q}, \lambda} \dfrac{1}{\omega_k\omega_q}  \int_{0}^{t} dt_1  \int_{0}^{t} dt_2 \int_{0}^{t_2} dt_3  \int_{0}^{t_3} dt_4 e^{-i\omega_k(t_1 - t_2)}  \nonumber \\
	&\qquad \bigg\{ e^{ -i\omega_q(t_3 - t_4)} \cdot 
	\langle [\hat{p}_{sc, \lambda}(t_3), \hat{p}_{sc, \sigma}(t_2)] \rangle \langle \hat{p}_{sc, \sigma}(t_1)\rangle \langle \hat{p}_{sc, \lambda}(t_4) \rangle \nonumber\\
	&\qquad -e^{ -i\omega_q(t_4 - t_3)} \cdot 
	\langle [\hat{p}_{sc, \lambda}(t_3), \hat{p}_{sc, \sigma}(t_2)] \rangle \langle \hat{p}_{sc, \sigma}(t_1)\rangle \langle \hat{p}_{sc, \lambda}(t_4) \rangle \bigg\} \nonumber\\
	&=\sum_{\bm{q}, \lambda} \dfrac{1}{\omega_k\omega_q}  \int_{0}^{t} dt_1  \int_{0}^{t} dt_2 \int_{0}^{t_2} dt_3  \int_{0}^{t_3} dt_4 e^{-i\omega_k(t_1 - t_2)}  \nonumber \\
	&\qquad 2\text{Re}\bigg\{ e^{ -i\omega_q(t_3 - t_4)} \cdot 
	\langle [\hat{p}_{sc, \lambda}(t_3), \hat{p}_{sc, \sigma}(t_2)] \rangle \langle \hat{p}_{sc, \sigma}(t_1)\rangle \langle \hat{p}_{sc, \lambda}(t_4) \rangle \bigg\}, 
	\label{eq:spec_highest_order_correction}
\end{align}
where we have used that $\int_0^t dt_1 \int_0^t dt_2 f(t_1, t_2) = \int_0^t dt_1 \int_0^{t_1} dt_2 f(t_1, t_2) + \int_0^t dt_1 \int_0^{t_1} dt_2 f(t_2, t_1)$. Finally we have found,
\begin{equation}
\langle \hat{a}_{\bm{k}, \sigma}'^\dagger (t) \hat{a}_{\bm{k}, \sigma}' (t) \rangle_{cor} 
= 	
g_0^4 N^3 \cdot 2\text{Re}\bigg[\langle \hat{T}^{(1)\dagger}_{\bm{k}, \sigma}(t) \hat{T}^{(3a)}_{\bm{k}, \sigma}(t) \rangle_{N^3} 
+\langle \hat{T}^{(1)\dagger}_{\bm{k}, \sigma}(t) \hat{T}^{(3b)}_{\bm{k}, \sigma}(t) \rangle_{N^3}  \bigg] + \mathcal{O}(g_0^4 N^2),
\label{eq:counting_operator_correction}
\end{equation}
where the expression in square brackets is given by Eq. (\ref{eq:spec_highest_order_correction}). We see that this highest order correction to the photonic counting operator does not obey the selection rules that determine the dominating coherent part of the spectrum in Eq. (\ref{eq:spectrum_perturbative_coherent}) as Eq. (\ref{eq:spec_highest_order_correction}) is not proportional to the norm-squared of the Fourier-transformed dipole which is the basis of the symmetry-bases selection rules for the harmonics spectrum \cite{Neufeld2019}.

The proper way to numerically evaluate Eq. (\ref{eq:spec_highest_order_correction}) remains an issue for future research. The expression is made complicated by the sum over all modes of the light, whose resolution and termination point are not well defined nor understood, making the evaluation uncontrolled. 

	\section{Product Ansatz in the Schrödinger picture} \label{App:product_ansatz_consequence}
	In this appendix, we show how the product ansatz used in the Schrödinger picture has unintended consequences, e.g., yielding only the coherent part of the spectrum, which is discussed in Sec. \ref{Sec:Dis_AdvantagesPHD} in the main text.
	In the Schrödinger picture, treating the full bosonic Hilbert space of all photonic modes is numerically prohibitive, and several approximations are therefore introduced. A common one is to treat all photonic modes independently, i.e., to neglect coupling between frequency components~\cite{Gorlach2020, Lange2024a, Lange2025a, Lange2025b}. The total state is written as
	\begin{equation}
		\ket{\Psi(t)} = \sum_m \ket{\phi_m}\ket{\chi^{(m)}(t)},
		\label{eq:Born_Huang}
	\end{equation}
	where $\ket{\phi_m}$ is the $m$-th electronic eigenstate and $\ket{\chi^{(m)}(t)}$ its correlated photonic state. The decoupling approximation assumes a product structure,
	\begin{equation}
		\ket{\chi^{(m)}(t)} = \otimes_{\bm{k},\sigma} \ket{\chi_{\bm{k},\sigma}^{(m)}(t)},
		\label{eq:decoupling_ansatz}
	\end{equation}
	whose dynamics in each mode obey
	\begin{equation}
		i\partial_t \ket{\chi^{(m)}_{\bm{k},\sigma}(t)}
		= \hat{\bm{A}}_{\bm{k},\sigma}(t)\cdot 
		\sum_n \bm{p}_{m,n}(t)\ket{\chi^{(n)}_{\bm{k},\sigma}(t)},
		\label{eq:chi_eom}
	\end{equation}
	with $\hat{\bm{A}}_{\bm{k},\sigma}(t)= g_0/\sqrt{\omega_k}\,(\hat{a}_{\bm{k},\sigma}e^{-i\omega_k t}+{\rm h.c.})\,\hat{\bm{e}}_\sigma$ acting only on mode $(\bm{k},\sigma)$ and initial condition $\ket{\chi_{\bm{k}, \sigma}^{(m)}} = \delta_{m,i} \ket{0}_{\bm{k}, \sigma}$. Thus, Eq.~(\ref{eq:chi_eom}) contains no coupling between modes.
	
	Expanding $\ket{\chi^{(m)}_{\bm{k},\sigma}}$ in a Fock basis and evaluating expectation values using Eq.~(\ref{eq:Born_Huang}) yields~\cite{Lange2024a}
	\begin{align}
		\langle \hat{a}^\dagger_{\bm{k}',\sigma'} \hat{a}_{\bm{k}',\sigma'} \rangle
		&= \sum_m 
		\bra{\chi_{\bm{k}',\sigma'}^{(m)}} 
		\hat{a}^\dagger_{\bm{k}',\sigma'}\hat{a}_{\bm{k}',\sigma'} 
		\ket{\chi_{\bm{k}',\sigma'}^{(m)}} 		\prod_{\substack{\bm{k},\sigma\\\neq(\bm{k}',\sigma')}} 
		\langle\chi_{\bm{k},\sigma}^{(m)}(t)\vert\chi_{\bm{k},\sigma}^{(m)}(t)\rangle,
		\label{eq:expectation_value_scrodinger_picture}
	\end{align}
	illustrating that expectation values in one mode depend on overlaps in all other modes, even though their time evolution is independent. This discrepancy is an artifact of the product (decoupling) ansatz.
	
	Numerically, the product over other modes behaves approximately as a Kronecker delta in $m$, i.e.,
	\[
	\prod_{\substack{\bm{k},\sigma\\\neq(\bm{k}',\sigma')}} 
	\langle\chi_{\bm{k},\sigma}^{(m)}\vert\chi_{\bm{k},\sigma}^{(m)}\rangle
	\simeq \delta_{m,i},
	\]
	with $i$ the initial electronic state. This behaviour can be motivated analytically by integrating Eq.~(\ref{eq:chi_eom}) to first order:
	\begin{align}
		\ket{\chi_{\bm{k},\sigma}^{(m)}(t)}^{(1)}
		&= \delta_{m,i}\ket{0}_{\bm{k},\sigma} - i\frac{g_0}{\sqrt{\omega_k}}
		\int_0^t dt'\,
		\hat{\bm{e}}_\sigma\cdot\bm{p}_{m,i}(t')\,e^{i\omega_k t'}
		\ket{1}_{\bm{k},\sigma},
	\end{align}
	leading to overlaps
	\begin{align}
		&\prod_{\substack{\bm{k},\sigma\\\neq(\bm{k}',\sigma')}} \! \! \!~^{(1)}\langle\chi_{\bm{k},\sigma}^{(m)}(t)\vert\chi_{\bm{k},\sigma}^{(m)}(t)\rangle^{(1)} = \prod_{\substack{\bm{k},\sigma\\\neq(\bm{k}',\sigma')}} 
		\Big[
		\delta_{m,i}
		+ \frac{g_0^2}{\omega_k}
		\big|\!\int_0^t dt'\,e^{i\omega_k t'}\hat{\bm{e}}_\sigma\!\cdot\!\bm{p}_{m,i}(t')\big|^2
		\Big],
		\label{eq:overlap_final}
	\end{align}
	whose first term dominates when calculating the product since $g_0^2/\omega_k \ll 1$. Equation~(\ref{eq:overlap_final}) explains why, within the decoupling ansatz, only the photonic state $\ket{\chi_{\bm{k},\sigma}^{(i)}(t)}$ contributes to expectation values, and consequently why the Schrödinger-picture treatments of Refs.~\cite{Lange2024a, Lange2025a, Lange2025b} capture only the coherent part of the spectrum.
	
	\twocolumngrid
	\bibliography{big_bibliography}
	
\end{document}